\documentclass[iop]{emulateapj}
\usepackage{amssymb,fancyhdr,lastpage,layout,bold-extra,floatrow,calc,wrapfig}
\usepackage[backref,breaklinks,colorlinks,allcolors=blue]{hyperref}
\usepackage[all]{hypcap}
\usepackage{graphicx}
\usepackage{subfigure}
\usepackage{natbib}
\usepackage{pdfpages}
\usepackage{amsmath}
\usepackage{mathtools}
\usepackage{setspace}
\usepackage{xspace}
\usepackage{float}
\usepackage{relsize}
\newcommand{\lir}{L_{\rm IR}}

\newcommand{\mdust}{M_{\rm dust}}

\newcommand{\msun}{~\mathrm{M_{\odot}}}
\newcommand{\lsun}{~\mathrm{L_{\odot}}}
\newcommand{\sed}{\lambda L_{\lambda}}
\newcommand{\gadgettwo}{\textsc{gadget-2}\xspace}
\newcommand{\sunrise}{\textsc{sunrise}\xspace}

\begin{document}

\title{What shapes the far-infrared spectral energy distributions of galaxies?}

\author{Mohammadtaher Safarzadeh\altaffilmark{1}, Christopher C. Hayward \altaffilmark{2,3,$\dagger$},
Henry C. Ferguson\altaffilmark{4}, Rachel S. Somerville\altaffilmark{5}}

\shorttitle{What shapes the FIR SEDs of galaxies?}
\shortauthors{M.~Safarzadeh et al.}

\altaffiltext{1}{Johns Hopkins University, Department of Physics and
Astronomy, 366 Bloomberg Center, 3400 N. Charles Street, Baltimore, MD
21218, USA, Email: mts@pha.jhu.edu}
\altaffiltext{2}{TAPIR 350-17, California Institute of Technology, 1200 E. California Boulevard, Pasadena, CA 91125, USA}
\altaffiltext{3}{Harvard--Smithsonian Center for Astrophysics, 60 Garden Street, Cambridge, MA 02138, USA}
\altaffiltext{4}{Space Telescope Science Institute, 3700 San Martin Boulevard, Baltimore, MD 21218, USA}
\altaffiltext{5}{Department of Physics and Astronomy, Rutgers, The State University of New Jersey, 136 Frelinghuysen Road, Piscataway, NJ 08854, USA}
\altaffiltext{$\dagger$}{Moore Prize Postdoctoral Scholar in Theoretical \\ Astrophysics}

\begin{abstract}
To explore the connection between the global physical properties of galaxies and 
their far-infrared (FIR) spectral energy distributions (SEDs),
we study the variation in the FIR SEDs of a set of hydrodynamically simulated galaxies
that are generated by performing dust radiative transfer in post-processing.
Our sample includes both isolated and merging systems at various stages of the merging process and covers 
infrared (IR) luminosities and dust masses that are representative of both low- and high-redshift galaxies.
We study the FIR SEDs using principle component analysis (PCA) and find that 97\% of the variance
in the sample can be explained by two principle components (PCs). The first PC characterizes the wavelength of
the peak of the FIR SED, and the second encodes the breadth of the SED.
We find that the coefficients of both PCs can be predicted well using a double power law in terms of the IR
luminosity and dust mass, which suggests that these two physical properties are the primary determinants of
galaxies' FIR SED shapes. Incorporating galaxy sizes does not significantly improve our ability to predict the FIR SEDs.
Our results suggest that the observed redshift evolution in the effective dust temperature at fixed IR luminosity
is not driven by geometry: the SEDs of $z \sim 2-3$ ultraluminous IR galaxies (ULIRGs) are cooler than those of local ULIRGs
not because the high-redshift galaxies are more extended but rather because they have higher dust masses at fixed IR
luminosity. Finally, based on our simulations, we introduce a two-parameter set of SED templates that depend on
both IR luminosity and dust mass.
\end{abstract}

\section{Introduction} \label{S:intro}

Understanding what drives the variation in the far-infrared (FIR) spectral energy distributions (SEDs) of galaxies is
a key goal if we wish to maximize the physical insight that can be gleaned from the wealth of data that is rapidly
being collected in the era of \emph{Herschel} \citep{Pilbratt2010} and the Atacama Large Millimeter Array (ALMA).
Unfortunately, interpreting FIR SEDs is difficult because spatially resolved data are not available for the vast
majority of galaxies (although ALMA will help greatly in this regard). Moreover, there are various degeneracies (e.g., between
the dust temperature distribution and the power-law index of the dust emissivity curve; see \citealt{Kelly:2012} and references
therein) that are difficult to break.

FIR SEDs are fit using a wide variety of methods (see \citealt{casey14} for a recent review).
Often, one or a sum of a few modified blackbody SEDs are used to fit FIR SEDs, but the
physical information that can be gained from such models is limited (see, e.g., \citealt{H12} and \citealt{Smith:2013} for detailed
discussions). Another common approach is to use templates: by fitting the available photometry with an IR SED template, the IR
luminosity can be estimated. Sometimes one or a few SEDs of well-studied IR luminous galaxies, such as Arp 220, are
used. Various empirically based, single-parameter templates are also widely used \citep[e.g.,][]{CE01,Pope2008,R09,Magdis2012}.

Although empirically based FIR SED templates have many practical uses, physical models are necessary if one wishes to learn
about, e.g., the radiation field and dust properties of a galaxy.
Because of the computational expense of radiative transfer and difficulty of constraining many free parameters with a limited number of FIR photometric data points,
approaches to modeling FIR SEDs without doing radiative transfer calculations have been advocated
\citep[e.g.,][]{Desert1990,Devriendt1999,2001Dale,2002DH,draine_li_07,magphys,2012Somerville}.
Some authors \citep[e.g.,][]{2001Dale,2002DH} model FIR SEDs by assuming that the mass distribution of dust that is exposed 
to radiation fields of different intensities can be described by a truncated power law. \citet{draine_li_07} expanded this model by adding
a delta function to the intensity distribution (at the minimum intensity) to represent diffuse ISM dust (see also \citealt{Draine2007}).
Such models can provide good fits to observed SEDs. Similarly, other authors \citep[e.g.,][]{Kovacs:2010,Magnelli:2012} parameterize
SEDs by assuming a power-law distribution of dust temperatures.

Radiative transfer calculations, either with simplified assumptions about the 
geometry of dust with respect to stars \citep{1996aWitt,2000bWitt,1998Silva,2000Efstathiou,2001Gordon,2003Takagi,DeGeyter2014} or active
galactic nuclei (AGN; e.g., \citealt{Fritz2006,2007Siebenmorgen,Stalevski2012}),
or more complicated geometries \citep{2005Dopita,2011Popescu,DeLooze2012,DeLooze2014,DeGeyter2015},
have been qualitatively successful in producing SEDs similar to those observed.
It is also possible and (has become increasingly common) to `forward-model' FIR SEDs by performing three-dimensional (3D) dust radiative transfer
on the outputs of hydrodynamical simulations of galaxies in post-processing
\citep[e.g.][]{Jonsson06,Jonsson06b,Jonsson10,Chakrabarti:2007,Chakrabarti:2008,Chakrabarti:2009,N10,N10b,H11,H12,H14,Snyder:2013,GRASIL3D,
Lanz14,Martinez14,Granato15}.

Because explicit radiative transfer calculations can best capture complicated source and dust geometries; directly treat dust absorption (including not only absorption of
primary radiation from stars and AGN but also dust self-absorption), scattering, and re-emission; and fully characterize the 3D distribution of dust
temperature, which depend on both the local radiation field and the grain properties, they provide the best tool for studying how FIR SEDs depend
on galaxy properties.
By performing 3-D Monte Carlo radiative transfer calculations for idealized geometries (i.e., not outputs of hydrodynamical simulations),
\citet{2001Misselt} studied the dependence of the shape of the FIR SED on global parameters of the emitting regions.
They found that higher dust mass leads to colder SEDs when the other parameters in their model are fixed. This is a simple
consequence of thermal equilibrium. Moreover, a `shell' (aka foreground screen) geometry results in a broader SEDs compared
to a geometry in which stars and dust are mixed (which they term the `dusty' geometry) because of differences in the temperature structures of the two models.

Performing radiative transfer on hydrodynamical simulations of galaxy formation has the benefit of including more realistic source and
dust geometries than simpler approaches, such as that of \citet{2001Misselt}.
This approach has been demonstrated \citep{Jonsson10} to yield SEDs that agree well with
observed local samples of normal star forming galaxies from the Spitzer Infrared Nearby Galaxies Survey
\citep[SINGS;][]{2003Kennicutt} and local luminous infrared (IR) galaxies (LIRGs) from the Great Observatories All-sky LIRG Survey
\citep[GOALS;][]{2009Armus}.
Moreover, \citet{Lanz14} have shown that the SEDs of local interacting galaxies from the \emph{Spitzer} Interacting Galaxies Survey
\citep[SIGS;][]{Lanz2013,Brassington2015} can be fit
reasonably well with SEDs predicted in this manner; this is also the case for 24-$\mu$m-selected galaxies at $z \sim 0.3 - 2.8$
(Roebuck et al., in preparation). Unfortunately, performing 3D radiative transfer on hydrodynamical simulations is orders-of-magnitude more computationally expensive
than less detailed approaches that assume smooth axisymmetric geometries \citep[e.g.,][]{1998Silva}: the former typically requires of order $10-10^3$ CPU-hours
per galaxy, whereas the latter requires at most a few minutes. Thus, it is instructive to determine whether the high computational expense of
performing 3D dust radiative transfer on hydrodynamical simulations of galaxies can be avoided, as it may be possible to characterize the variation among the simulated SEDs
using only a few global parameters.

In this work, we investigate whether we can predict the FIR SEDs of simulated galaxies, which were calculated using 3D dust radiative transfer in previous works,
using a simple, computationally inexpensive method. The method that we use to study these SEDs is principle component analysis (PCA).
In our case, PCA yields FIR SED eigenvectors (principle components, hereafter PCs) such that linearly combining them with the mean SED of the 
sample can capture the variance in the simulated SEDs. The coefficients of each PC are different for each galaxy,
and studying how the coefficients depend on global parameters, such as the star formation rate (SFR) or IR luminosity, can give physical
intuition regarding what drives the dispersion in the FIR SEDs.

The simulated galaxy SEDs used in this work have been analyzed extensively in previous works. These or similar simulations have
been demonstrated to exhibit good agreement with the SEDs/colors of various classes of real galaxies, including local `normal' galaxies \citep{Jonsson10}
and (U)LIRGs \citep{Younger09,Jonsson10,Lanz14}, high-redshift dusty star-forming galaxies (\citealt{N10,N10b,H11,H12}; Roebuck et al., in prep.),
obscured AGN (\citealt{Snyder:2013}; Roebuck et al., in preparation), post-starburst galaxies \citep{Snyder11}, and
compact quiescent galaxies \citep{Wuyts10}. Thus, although the simulations used herein naturally have some associated limitations (see Section \ref{S:limitations}),
they have the advantage of being well-tested and in agreement with many observational constraints. Furthermore, to the best of our
knowledge, they still represent the state of the art in terms of ultraviolet (UV)--millimeter (mm) SEDs generated from 3D hydrodynamical simulations of galaxies.

In addition to determining what physical insights about galaxies can be gained from
their FIR SEDs, we hope to identify possibilities for improving
how semi-analytic models (SAMs) of galaxy formation predict FIR SEDs. Some SAMs
\citep[e.g.,][]{1998Silva,Granato:2000} employ radiative transfer calculations that assume axisymmetric geometries or analytic models
that are designed to capture the results of such calculations \citep[e.g.,][]{Gonzalez:2011}. Such calculations have been widely employed,
but it is unclear how well their results agree with those of 3D radiative transfer calculations performed on hydrodynamical simulations
of galaxies with more complex geometries.
Other SAMs \citep[e.g.,][]{2012Somerville}
use empirically derived templates that are a function of a single parameter, $\lir$, to predict FIR SEDs. However, such an approach
is insufficient; for example, it cannot capture the redshift evolution of the effective dust temperature--IR luminosity relation
(\citealt{casey14} and references therein) by construction. Thus, we aim to determine one or more additional
physical parameters, besides
$\lir$, that can be used to predict the FIR SEDs of galaxies. Having determined what additional parameter(s) is necessary to
predict the variation in IR SEDs, we can then generate a set of multi-parameter SED templates. By incorporating these templates in a
SAM, we may be able to resolve the tension between the observed FIR
and submillimeter number counts and those predicted by the SAM
\citep{2012Somerville,2012Niemi}; this will be explored in a future
work. These templates could be also used in semi-empirical models
\citep[e.g.,][]{Bethermin2012,Bethermin2013,Bernhard2014}. Finally,
these two-parameter templates will be useful for fitting observed SEDs
to infer the total IR luminosity and predict fluxes at wavelengths for
which data are not available.

The remainder of this paper is organized as follows: in Section \ref{sec:simulations}, we describe the properties of the simulated galaxy SED
dataset that we use and summarize the details of the hydrodynamical simulation and dust radiative transfer calculation methods.
Section \ref{sec:pca_method} summarizes the PCA technique and how we use the PCA results to predict the simulated galaxies' SEDs.
In Section \ref{sec:PCA}, we present the results of the PCA of the simulated galaxy SEDs. Section \ref{sec:dust_mass} demonstrates that
the dust mass is a key parameter, in addition to the IR luminosity, that determines the SED shape, whereas in Section \ref{S:size_results},
we show that incorporating the galaxy size does not improve our ability to predict the FIR SEDs. In Section \ref{S:templates}, we present
a two-parameter family of SED templates that depend on the IR luminosity and dust mass. In Section \ref{S:discussion}, we discuss some
observational support for the importance of dust mass in determining the SED shape; speculate regarding the unimportance of galaxy
size, the origin of catastrophic failures to predict the SEDs, and the possible implications of using the proposed two-parameter
templates in SAMs; and highlight some limitations of our work. Section \ref{S:conclusions} lists our primary conclusions.

\section{Simulated galaxy SED dataset}\label{sec:simulations}

In this work, we analyze two sets of FIR SEDs of simulated isolated and merging disk galaxies that were presented in previous works. We will
first summarize the properties of the two datasets and then briefly discuss the simulation methodology.
The first SED dataset was originally presented in \citet{Lanz14} and \citet{H14}. The four progenitor galaxies span a stellar (baryonic) mass range of
$6 \times 10^8 - 4 \times 10^{10} \msun$ ($10^9 - 5 \times 10^{10} \msun$), and their properties were selected to be representative of typical star-forming
galaxies in the $z \sim 0$ universe
(see \citealt{Cox08} for details). Each of the four progenitors was simulated in isolation, and binary mergers of all possible combinations of progenitors
were simulated for a single representative orbit (i.e., the results for other `non-special' orbits are similar; the results for perfectly co-planar and other rare
special configurations can sometimes differ significantly; e.g., \citealt{Cox08}), thereby yielding 10 merger
simulations. The total dataset contains $\sim 12,000$ SEDs. We refer to this dataset as the $z \sim 0$ dataset.

The second set of SEDs of simulated isolated disk and merging galaxies was presented in \citet{H11,H12,H13}.
For this dataset, the structural properties of the progenitor (bulgeless) disk
galaxies were scaled to $z \sim 3$ following the method of \citet{Robertson06}, and the initial gas fractions of the disks, 0.6-0.8, are significantly greater
than those of the $z \sim 0$ simulations.\footnote{Such large gas fractions were used to ensure that the gas fraction at the time of coalescence was
consistent with observational constraints for $z \sim 2$ galaxies. For all snapshots considered in this work, the gas fractions are $< 0.5$. See \citet{H13}
for more details.}
Because the original purpose of these simulations was to model high-redshift submillimeter galaxies (SMGs),
the progenitor disks span a relatively narrow baryonic mass range of $\sim 1-4 \times 10^{11} \msun$, but a variety of merger orbits and mass ratios are included
(see \citealt{H12} for details). This dataset contains 37 hydrodynamical simulations, from which $\sim 6500$ SEDs were calculated.
We refer to this dataset as the $z \sim 3$ dataset.

The SEDs contained in the two datasets were generated by performing dust radiative transfer in post-processing on the outputs of 3-D hydrodynamical
simulations. The full methodology is described in the aforementioned works and references therein, so we will only summarize it here. First,
idealized isolated (i.e., non-cosmological) galaxy models were created following the method described in \citet{Springel05feedback}.
Each initial disk galaxy is composed of a dark matter halo, stellar and gaseous disks, and a supermassive black hole (SMBH); for the $z \sim 0$ simulations
only, a stellar bulge is also included. Then, the isolated galaxies and binary mergers of these galaxies were simulated using a heavily modified
version of the \gadgettwo $N$-body/smoothed-particle hydrodynamics (SPH) code \citep{Springel05gadget}.\footnote{The traditional density--entropy formulation
of SPH was used. However, we note that the results of idealized non-cosmological galaxy simulations such as these are relatively insensitive to the numerical
method \citep{H14arepo} and thus do not consider the use of traditional SPH to be a significant limitation.}
The simulations include the effects of gravity, hydrodynamical interactions, and radiative heating and cooling. Star formation and stellar feedback
are incorporated via the two-phase sub-resolution interstellar medium (ISM) model of \citet{Springel03}, and BH accretion and AGN feedback are treated following
\citet{Springel05feedback}. Each gas particle is enriched with metals according to its associated SFR, assuming a yield of 0.02.

Every 10 Myr, `snapshots' of the physical state of the simulations were saved. Then, to calculate UV--mm SEDs, 3-D dust radiative transfer
was performed in post-processing on a subset of the snapshots using the code \sunrise \citep{Jonsson06,Jonsson10}. For a given snapshot, the
\sunrise calculation proceeds as follows: the stellar and BH particles in the \gadgettwo simulation, which are the sources of radiation,
are assigned source SEDs according to their properties (age and metallicity for the star particles and luminosity for the BH particles). The metal
distribution from the simulation is projected onto an octree grid for the purpose of calculating the dust optical depths (assuming a dust-to-metal
density ratio of 0.4; e.g., \citealt{Dwek:1998,James:2002,Sparre:2014}). The Milky Way $R_V = 3.1$ dust model of \citet{draine_li_07} was used.

The most significant uncertainty (of which we are aware) in these calculations is the sub-resolution structure of the ISM.
Specifically, the simulations do not resolve the ISM structure on scales of less than a few hundred parsecs, but observations clearly indicate that
the ISM has substantial structure on smaller scales. We can crudely investigate this uncertainty through the use of two alternate treatments for
the sub-resolution ISM.\footnote{This issue has been discussed extensively in previous works \citep[e.g.,][]{H11,Snyder:2013,Lanz14,HS15},
and we refer the interested reader to those papers for additional details.}
In the first treatment (referred to as `multiphase on' or `default ISM' in previous works), it is assumed that the cold clouds implicit in the
\citet{Springel03} model have negligible volume filling factors. Thus, for the purpose of the radiative transfer calculations, this dust is ignored.
To account for the obscuration of young stars from their birth clouds, a sub-resolution model for HII and photodissociation regions \citep{Groves:2008} is
used. In the second treatment (`multiphase off' or `alternate ISM'), the entire dust mass (i.e., dust in both the implicit diffuse phase and cold clouds) is considered
in the radiative transfer calculations. In this work, we use the `multiphase off' model to ensure that all of the dust emission is calculated self-consistently (i.e., there
is no component from the sub-resolution HII and photodissociation region model). However, we have checked that our conclusions are qualitatively
unchanged if the `multiphase on' model is used.

After the source and dust properties are specified, radiation transfer is performed using the Monte Carlo method to calculate the effects of dust absorption,
scattering, and re-emission. Importantly, dust self-absorption is treated using an iterative procedure in which the transfer of IR radiation and dust
temperature calculation are repeated until the luminosity absorbed in each cell is sufficiently converged.
For each snapshot, this process yields spatially resolved UV--mm SEDs of the simulated galaxy/merger viewed
from 7 viewing angles.

We only analyze the SEDs of galaxies that are either isolated systems or mergers that are experiencing coalescence
(defined by a black hole separation of \textless1 kpc) or are post-coalescence. The reason for this choice is that we only want to study systems that
can be considered a single galaxy. Using this black hole separation criterion excludes early stage mergers, which would often be considered
separate systems in low-redshift observations (although they would be unresolved in the FIR at high redshift when observed with single-dish FIR
or submillimeter telescopes). Moreover, when the system consists
of widely separated galaxies, the radiative transfer within individual galaxies is decoupled (i.e., the radiation within one galaxy does not have
a significant effect on the dust in the other galaxy when the two galaxies are separated by many kiloparsecs). Using a less conservative
criterion, such as 5 kpc, does not qualitatively affect our results.

\begin{figure}
\centering
\vskip -0.0cm
\resizebox{3.0in}{!}{\includegraphics[angle=0]{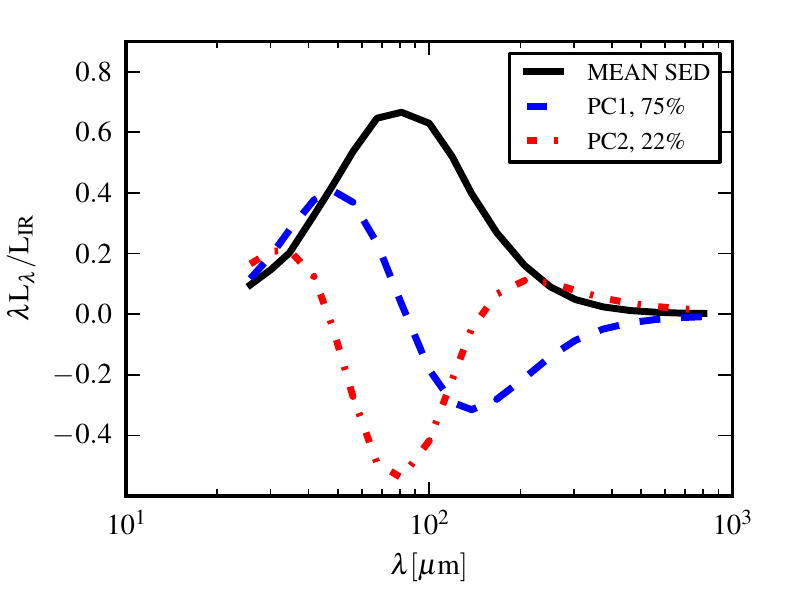}}
\caption{The black solid line corresponds to the mean FIR SED of our full sample; it has been normalized by dividing by $\lir$.
The first (PC1; blue dashed line) and second (PC2; green dash-dotted line) PCs are also shown.
These PCs are added to the mean with a unique coefficient for each galaxy to reconstruct the individual FIR SEDs.
The fraction of the variance in the data that is explained by each PC is indicated as a percentage in the legend.
Together, the top 2 PCs can explain 97\% of the total variance of the data in the full SED sample, which includes both low- and high-redshift
simulated galaxies.}
\label{fig:PC_all}
\end{figure}

\section{Predicting the SEDs based on PCA} \label{sec:pca_method}

Given a sample of FIR SEDs, PCA finds the mean SED of the entire sample ($\langle \sed \rangle$) and based on the mean SED, 
finds the PCs that encapsulate most of the variance of the entire sample. PCs are eigenvectors that are orthogonal to each other.
The PCs are not guaranteed to have a physical interpretation; rather, they are a tool
to reduce the dimensionality of the problem. The PCs are ranked in terms of how much of the variance is explained by each PC.

Because we are interested in studying the shape of the SEDs, we normalize the FIR SEDs by dividing them by the
IR luminosity.\footnote{We define $\lir$ as the integral of the SED over the wavelength range of 8-1000 \micron.}
The units of the data on which the PCA is performed do not affect the results, and we have chosen to work in dimensionless
units of $\lambda L_{\lambda}/\lir$.
 
The SED belonging to a given galaxy ($j$) in the sample can be reconstructed as a linear combination of principle components:
\begin{equation}
\lambda L_{\lambda, j} = \langle \sed \rangle + \sum_{i=1}^{N} C_{i,j} \times PC_{i},
\end{equation}
where $N$ is the number of PCs used, $PC_{i}$ is the $i$th PC, and $C_{i,j}$ is the coefficient of the $i$th PC for the $j$th
galaxy.

Having identified the most important PCs, we then examine how the coefficients of the PCs correlate with various
global physical parameters of the galaxies. Our goal is to be able to predict the PC coefficients, and thus the FIR SED,
of a galaxy based on a small number of galaxy properties, such as the SFR. To predict the PC coefficients, we use
functions of the following form:
\begin{equation} \label{eq:predicted_coeffs}
C'_{i,j} = \alpha_i + \sum_{k=1}^M \beta_{i,k} \log P_{i,j,k},
\end{equation}
where $C'_{i,j}$ is the predicted value of the $i$th coefficient for galaxy $j$,
$\alpha_i$ is the fit intercept for the $i$th PC coefficient, $\beta_{i,k}$ is the fit coefficient for the $i$th PC coefficient
and $k$th parameter, and $P_{i,j,k}$ is galaxy $j$'s value for the $k$th physical property used to predict the $i$th PC coefficient;
example properties include the SFR and dust mass.
As we will discuss below, we found that 2 PCs were sufficient to explain 97\% of the variance. Thus, we used $N=2$.
To predict the PC coefficients, we tried relations with both a single physical property ($M=1$) and a pair of physical
properties ($M=2$).

\begin{figure}
\centering
\vskip -0.0cm
\resizebox{3.5in}{!}{\includegraphics[angle=0]{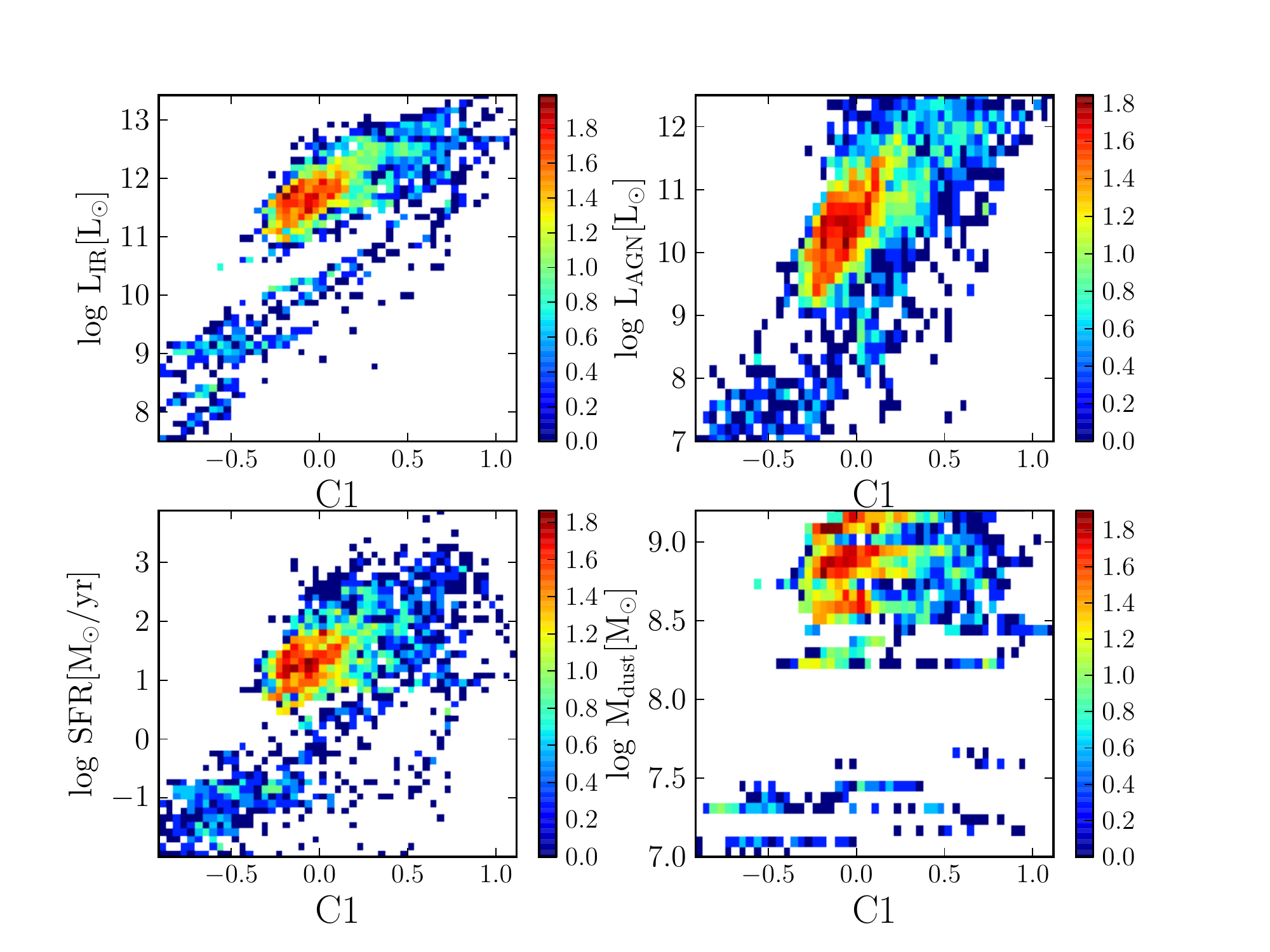}}
\resizebox{3.5in}{!}{\includegraphics[angle=0]{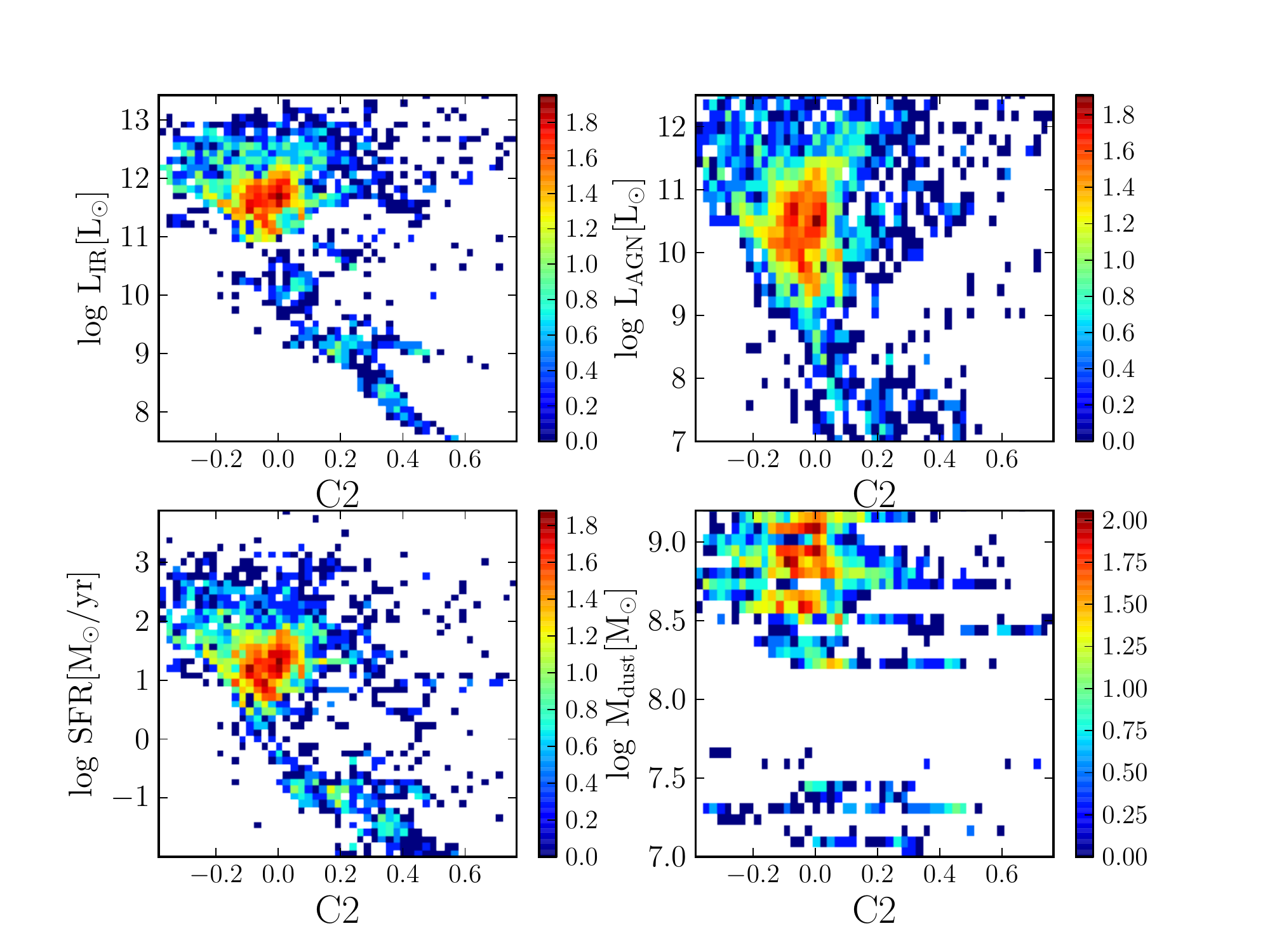}}
\caption{The \emph{top} set of four figures shows how the coefficient of the first PC depends on four global physical parameters of the galaxies
in the entire sample: IR luminosity (\emph{upper left}), AGN luminosity (\emph{upper right}), SFR (\emph{lower left}), and dust mass
(\emph{lower right}). The \emph{bottom} set of four figures shows the same plots for C2. In each panel, the parameter value is shown
on the $y$-axis, and the $x$-axis corresponds to the value of the coefficient.
The color of each bin corresponds to the logarithm of the number of points in the bin, as specified by the colorbars.
C1 is correlated with IR luminosity, SFR and AGN luminosity, although there is a large scatter. This result indicates that as the
IR luminosity, SFR or AGN luminosity is increased, the SED peak shifts toward shorter wavelengths. C2 is anti-correlated with
all of the four physical parameters, although the correlations are weaker (see the text for details). This result indicates that
there is a weak tendency for higher-luminosity or/and higher-dust mass galaxies to have narrower FIR SEDs.}
\label{fig:PC_correlation_all}
\end{figure}

Given predicted values for a galaxy's PC coefficients determined using Equation (\ref{eq:predicted_coeffs}), we predict
its SED as follows:
\begin{equation} \label{predicted_sed_eq}
\lambda L'_{\lambda,j} = \langle \lambda L_{\lambda} \rangle + \sum_{i=1}^{N} C'_{i,j}
\times PC_{i}.
\end{equation}

To quantify how well an SED can be predicted, we use the reduced chi squared value:
\begin{equation} \label{eq:chi_sq}
\chi^2_r = \frac{1}{df} \sum_{i=1}^P \frac{(L_{\lambda_i} - L'_{\lambda_i})^2}{\sigma_{\lambda_i}},
\end{equation}
where where $L_{\lambda_i}$ and $L'_{\lambda_i}$ denote the true and predicted luminosity density values at wavelength $\lambda_i$,
$\sigma_{\lambda_i}$ is the uncertainty at wavelength $\lambda_i$,
$P = 20$ is the total number of wavelength bins in the FIR, and
$df = P - M - 1$ is the number of degrees of freedom. Because our data are noise-free, for the purposes of calculating the
reduced $\chi^2$ value, we have arbitrarily assumed a signal-to-noise ratio of 5 in each band (i.e., $\sigma_{\lambda_i} = 0.2 L_{\lambda_i}$).
Thus, the $\chi_r^2$ values are useful in a relative sense (i.e., to determine which SEDs are predicted better than others), but the absolute values
are not very meaningful.

\section{PCA results}\label{sec:PCA}

We performed PCA on the FIR ($\lambda>25 ~\mu m$) SEDs from our entire sample, which includes both low- and high-redshift simulated galaxies. 
Figure~\ref{fig:PC_all} shows the mean SED and first two PCs of our sample. 
The percentage of the variance in the SEDs that can be explained by each PC is indicated 
in the figure. Each PC is multiplied by a coefficient (which is unique for each galaxy) and added 
to the mean SED of the sample to reconstruct the FIR SED of a particular galaxy. The coefficient can be negative or positive.
From Figure~\ref{fig:PC_all}, it is clear that adding the first PC (PC1) with a positive coefficient to the mean FIR SED tends
to make the SED warmer with respect to the mean SED (i.e., the wavelength at which the FIR emission peaks shifts to
shorter wavelengths); if it is added with a negative coefficient, the resulting SED is cooler
than the mean. Thus, the coefficient of PC1 can be considered a proxy for the effective dust temperature of the SED.
In contrast, the second PC (PC2) affects the width of the SED: if it is added with a positive coefficient to the mean FIR SED,
it tends to broaden the SED by increasing the power 
in the wings and removing power from the center. Conversely, it makes the FIR SED peak narrower if added with a negative coefficient. 

The coefficients of the PCs for a galaxy determine how its FIR SED differs from the mean SED. 
Figure~\ref{fig:PC_correlation_all} shows how the coefficients of PC1 and PC2 (we refer to them as C1 and C2, respectively) 
for each galaxy in the entire sample depend on four different
global physical parameters (SFR, AGN luminosity, IR luminosity and dust mass) that we expect to affect the FIR SEDs. We focus on these
specific parameters because they are simple global parameters that are clearly important for the radiative transfer, and they are typically available in
SAMs. The first two characterize the radiation that can potentially heat the dust. The IR luminosity tells us how much radiation is
absorbed by dust; of course, this quantity depends on both the bolometric luminosity of the stars and AGN and what fraction of
the intrinsic luminosity is absorbed.\footnote{Throughout this work, we use $\lir$ calculated by integrating the SEDs over the wavelength range
of $8-1000$ \micron. When dust self-absorption is negligible, this quantity is independent of viewing angle and identical to the luminosity
absorbed by dust. However, when dust self-absorption is non-negligible, $\lir$ can depend on viewing angle, whereas the luminosity
absorbed by dust is an intrinsic property of the simulated galaxy that does not depend on viewing angle (see \citealt{H11} for more details).
We opt to use $\lir$ rather than the absorbed luminosity because (1) the former can be inferred from observations without recourse to radiative
transfer modeling, (2) the average of $\lir$ taken over a sufficient number of viewing angles must equal the absorbed luminosity, and (3) dust
self-absorption does not lead to significant variation in $\lir$ with viewing angle for the bulk of the simulated galaxies.}
The dust mass characterizes the radiation sinks. As detailed below, we have also explored using other individual or pairs of parameters
to predict the coefficient values, but we do not show them in this plot because we do not focus on them in the bulk of the analysis below.

\begin{figure}
\centering
\vskip -0.0cm
\resizebox{3.0in}{!}{\includegraphics[angle=0]{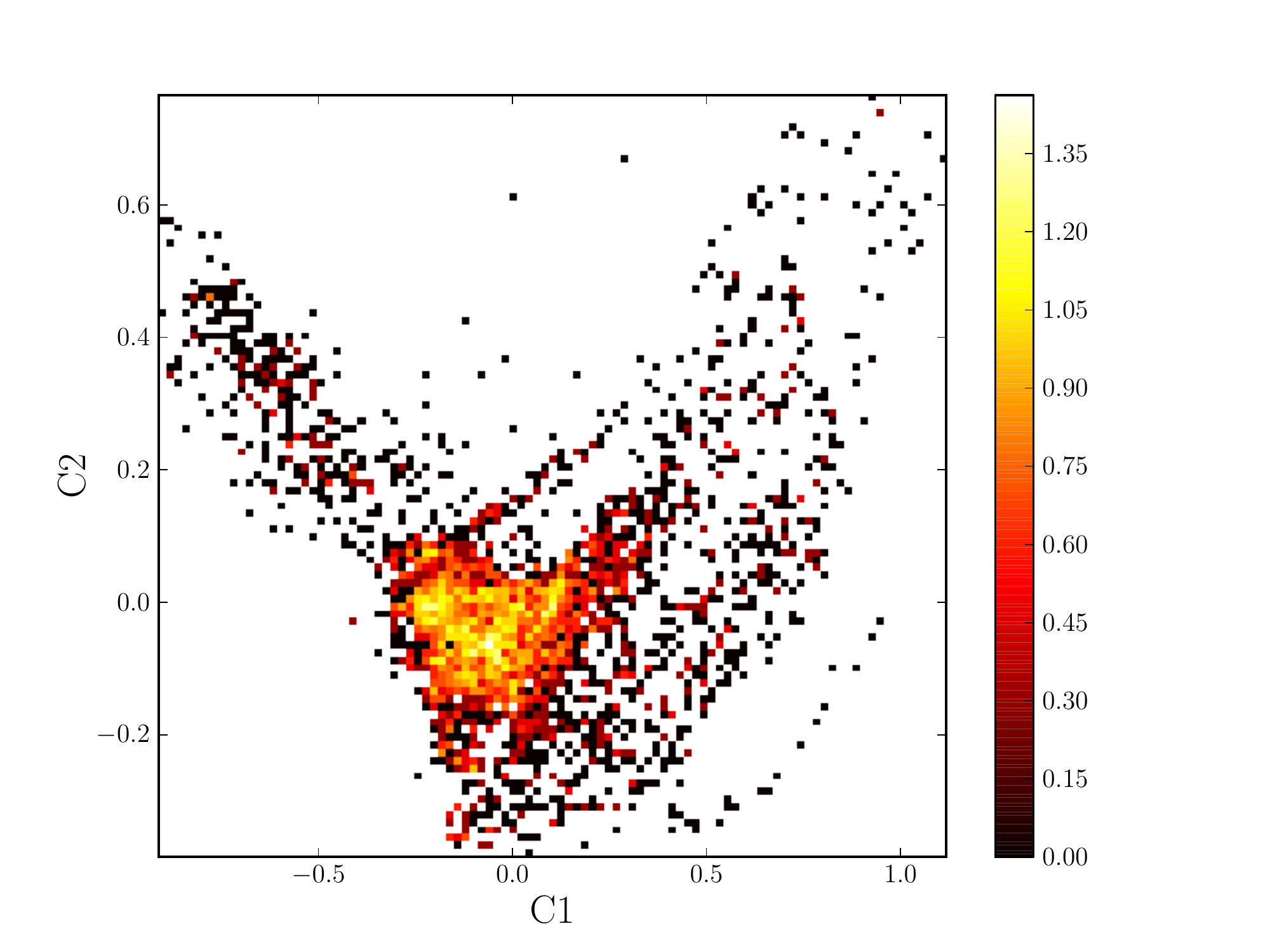}}
\caption{Density plot of C1 versus C2. The color bar indicates the logarithm of the number of SEDs in the bin.
Recall that higher values of C1 correspond to hotter SEDs, and higher values
of C2 correspond to broader SEDs. C1 and C2 are uncorrelated.}
\label{fig:C1_vs_C2}
\end{figure}

Figure \ref{fig:PC_correlation_all} indicates that C1 correlates with IR luminosity (with a Pearson correlation coefficient of $r = 0.71$),
SFR ($r = 0.60$) and AGN luminosity ($r = 0.73$), although there is a large scatter. This result indicates that as the
IR luminosity, SFR or AGN luminosity is increased, the SED peak shifts toward shorter wavelengths, which is to be expected because
of the known correlation between effective dust temperature and IR luminosity.
The correlations between C2 and the four properties shown in Figure \ref{fig:PC_correlation_all} are all weak negative correlations
(all have $-0.5 \la r < 0$). This result indicates that
there is a weak tendency for higher-luminosity or/and higher-dust mass galaxies to have narrower FIR SEDs.

\begin{figure}
\centering
\vskip -0.0cm
\resizebox{3.0in}{!}{\includegraphics[angle=0]{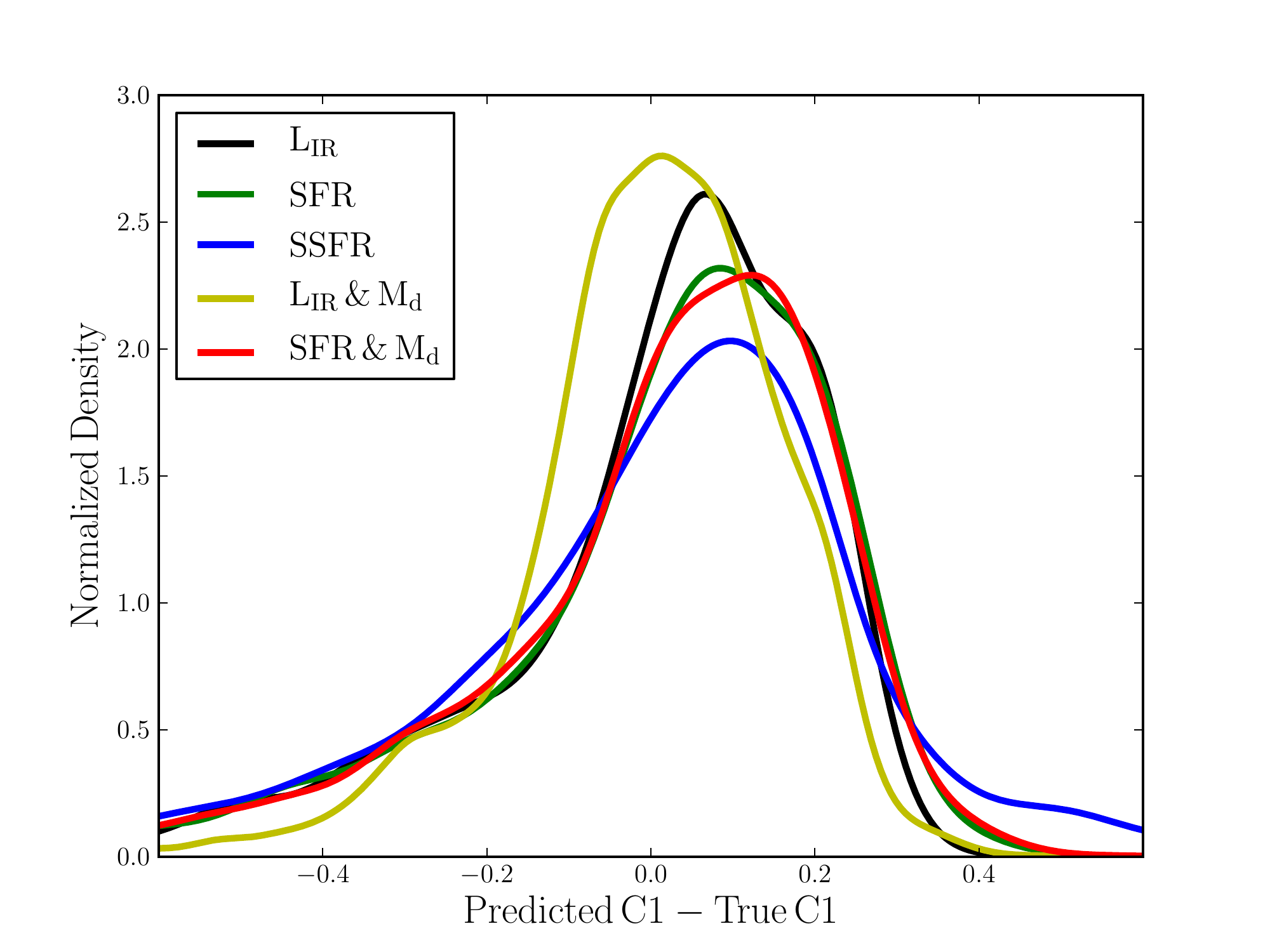}}
\resizebox{3.0in}{!}{\includegraphics[angle=0]{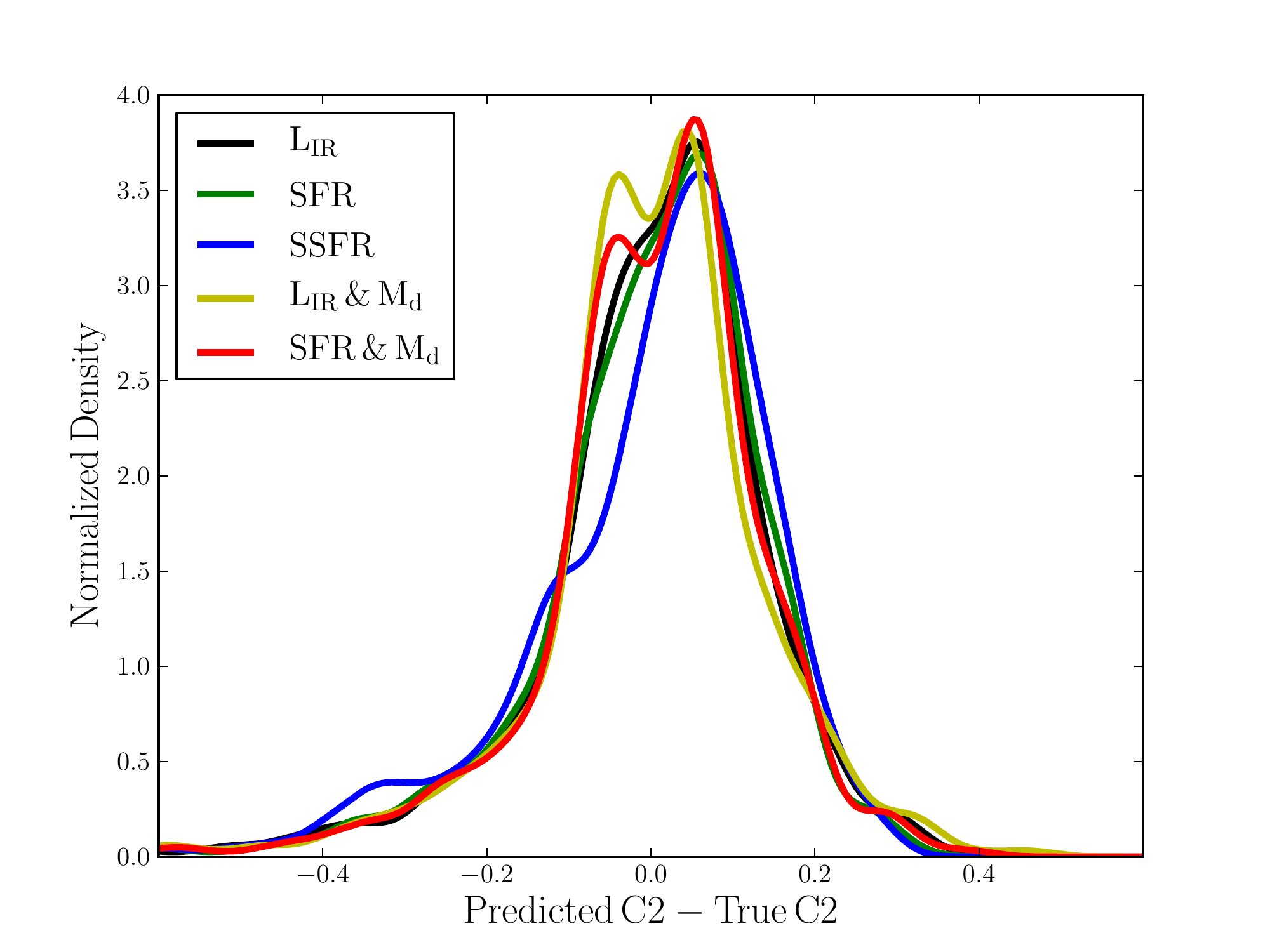}}
\caption{The \emph{top} panel shows the difference between the predicted and true values of C1 (the coefficient of the first PC).
C1 is estimated using five different estimators that depend on either one or two physical parameters; see the figure legend for the
parameters used. In all cases, the logarithm of the parameter is used. Using both IR luminosity and dust mass helps reduce the bias in the
predicted C1 values compared with the other estimators. In the \emph{bottom} panel, the same is shown for C2. Again, the combination
of IR luminosity and dust mass is superior to the others, although the difference in predictive power is less than for C1.}
\label{fig:hist_compare}
\end{figure}

Other than correlations between the PCs and the global parameters, it is also
interesting to consider whether there is a correlation between the PC coefficients. Although PC1 and PC2 are orthogonal by construction,
their coefficients may still be correlated.
In our case, a correlation between C1 and C2 would indicate a relation between the effective temperature of the SED and its broadness.
To see if there is such correlation, we plot C1 versus C2 for all the SEDs in our sample in Figure~\ref{fig:C1_vs_C2}.
We find that C1 and C2 are uncorrelated.

Because the scatter in the correlations between the coefficients and global galaxy parameters shown in Figure~\ref{fig:PC_correlation_all}
is large, it is worth considering whether we can better predict the coefficients
using two parameters simultaneously. In particular, thermal equilibrium considerations suggest that combining dust mass and a luminosity-related
parameter (e.g. IR luminosity or SFR) could be promising.\footnote{We have not used the AGN luminosity as a predictor despite the correlation
evident in Figure~\ref{fig:PC_correlation_all} for the following reason: for most of the simulated galaxies, the AGN contributes less than 10 percent
of the bolometric luminosity, and the FIR luminosity is not AGN-dominated (a detailed analysis of the contribution of AGN to FIR emission will be
presented in Hayward et al., in prep.) Consequently, the correlation between the AGN luminosity and C1 does not indicate causation; rather,
it arises because in the simulations that we analyze, the black hole accretion rate and SFR are correlated.}
Thus, in Figure~\ref{fig:hist_compare}, we show the results of predicting C1 (top) and C2 (bottom) in four
different manners: 1) using $\log \lir$ (e.g., C1 $= A \log \lir + B$), 2) using $\log$ SFR (e.g., C1 $= A \log$ SFR $+ B$), 3) using $\lir$ and $\mdust$
(e.g., C1 $= A \log \lir + B \log \mdust + C$) and 4) using $\log$ SFR together with $\log \mdust$ (e.g., C1 $= A \log$ SFR $+ B \log \mdust + C$).\footnote{We
have explored using various other combinations of parameters and found that these combinations had the best predictive power.}
It is clear that the combination of $\log \lir$ and $\log \mdust$ results is best able to predict C1, and this combination is superior to
$\log \lir$ alone (i.e., incorporating the dust mass reduces the error in the prediction).
For C2, the combination of $\log \lir$ and $\log \mdust$ is again superior to the others, but the differences in predictive power are less significant
than for C1. Our best-fitting relations for C1 and C2 are the following:
\begin{eqnarray}
{\rm C1} &=& 0.52 \log \left(\frac{\lir}{\lsun}\right) - 0.47 \log \left(\frac{\mdust}{\msun}\right) - 1.88 \label{eq:c1} \\
{\rm C2} &=& -0.014 \log \left(\frac{\lir}{\lsun}\right) - 0.10 \log \left(\frac{\mdust}{\msun}\right) \label{eq:c2} \\
& & + 1.05. \nonumber
\end{eqnarray}
We note that the coefficients for $\log \lir$ and $\log \mdust$ in Equation (\ref{eq:c1}) have similar magnitudes but opposite signs. This
suggests that the value of C1 depends on the ratio $\lir/\mdust$. We shall discuss this in detail in Section \ref{S:simplicity}. C2 depends
only weakly on dust mass and is effectively independent of $\lir$. Thus, there is a mild tendency for sources with higher dust masses
to have narrower SED peaks. Because of the weakness of the dependence, we will not interpret it further.

\begin{figure*}
\centering
\vskip -0.0cm
\resizebox{7.0in}{!}{\includegraphics[angle=0]{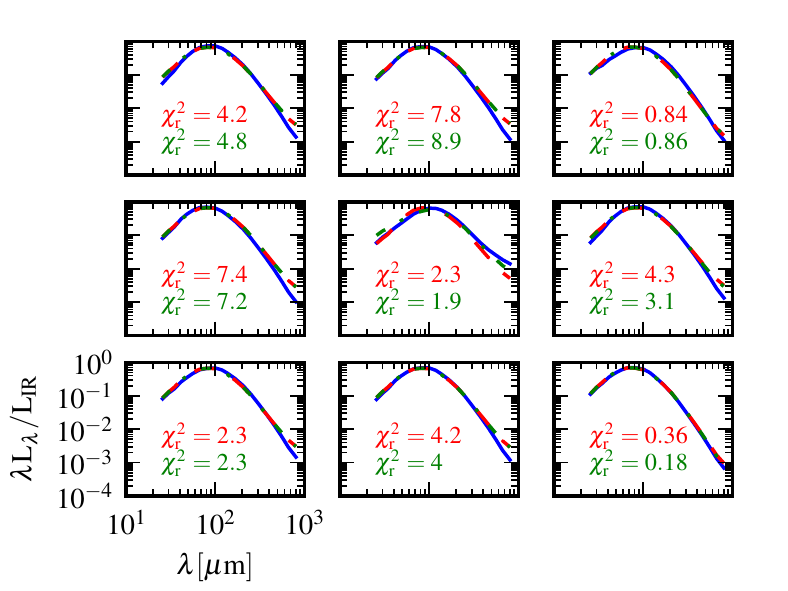}}
\caption{Comparisons of the predicted and true SEDs for 9 SEDs randomly chosen from the entire sample. The blue line is the actual SED.
The red (green) line is the predicted SED obtained by adding the first PC (first and second PC) to the mean SED
with the coefficient value(s) predicted using $\lir$ and $\mdust$. The numbers indicate the reduced $\chi^2$ value (Equation \ref{eq:chi_sq});
a signal-to-noise ratio of 5 in each band was arbitrarily assumed. Note that using the second PC leads to a better prediction in only some
of the cases.}
\label{fig:reconstruct}
\end{figure*}

\begin{figure}
\centering
\vskip -0.0cm
\resizebox{3.0in}{!}{\includegraphics[angle=0]{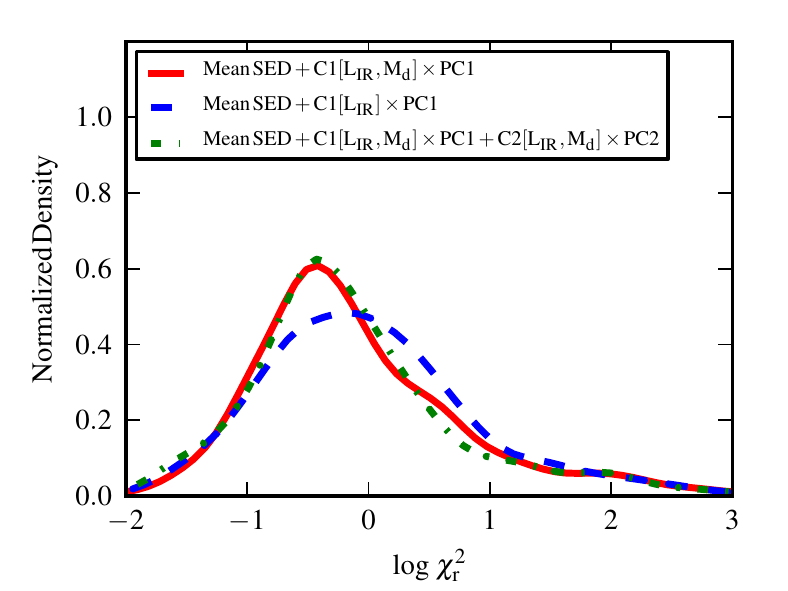}}
\caption{This figure shows how well the SEDs of the simulated galaxies can be predicted, as characterized by the $\chi^2_r$ values,
using the PCs. The red line corresponds to when only PC1 is used; its coefficient is predicted using $\lir$ and $\mdust$.
The dashed blue line shows the result when only PC1 is used and its coefficient is predicted using only $\lir$.
The dot-dashed green line shows the result when both PC1 and PC2 are used; their coefficients are predicted using
both $\lir$ and $\mdust$. When C1 is predicted using both $\lir$ and $\mdust$, the $\chi^2_r$ distribution shifts significantly to the left
compared with when only
$\lir$ is used; the median $\chi^2_r$ value is 0.56 (0.82) when both $\lir$ and $\mdust$ are (only $\lir$ is) used to predict C1.
Thus, incorporating the dust mass yields considerably better SED predictions. The similarity of the
red solid and green dot-dashed lines indicates that using PC2 does not lead to significantly better $\chi^2_r$ values, and
the median value (0.54) is only slightly less than when only PC1 is used (assuming that C1 is predicted using both $\lir$ and
$\mdust$).}
\label{fig:compare_C1_C2}
\end{figure}

We now investigate how well we can predict the SEDs by predicting the PC coefficients using Equations (\ref{eq:c1}) and
(\ref{eq:c2}) and using Equation (\ref{predicted_sed_eq}) to predict the SED. Figure \ref{fig:reconstruct} compares the predicted
and true SEDs for 9 randomly chosen SEDs from our entire sample.
The true SED is shown in blue. The red (green) line is the predicted SED obtained by adding the first PC (first and second PC)
to the mean SED with the coefficient value(s) predicted based on the galaxy's $\lir$ and $\mdust$ values (using Equations \ref{eq:c1}
and \ref{eq:c2}). The red (green) number in each panel indicates the reduced $\chi^2$ value (Equation \ref{eq:chi_sq}) obtained
when only PC1 (both PC1 and PC2) is (are) used to predict the SED. From these examples, we see that the FIR SEDs are generally
predicted very well near their peaks. Generally, when $\chi^2_r > 1$, the reason is that the predicted and actual FIR SEDs differ
at long wavelengths. Finally, using PC2 improves the prediction (in particular, at long wavelengths)
in only some cases, and it can actually make the prediction worse;
this is true even if we use the true values of C1 and C2 rather than those predicted using Equations (\ref{eq:c1}) and (\ref{eq:c2}).
We shall discuss these points in more detail below.

Figure~\ref{fig:compare_C1_C2} shows the distribution of $\chi^2_r$ (Equation \ref{eq:chi_sq}) for all SEDs in the full sample when the SEDs
are predicted in this manner (the green line). We also show the $\chi^2_r$ distributions obtained when using only PC1 and predicting its coefficient,
C1, using either $\lir$ alone (the blue dashed line) or $\lir$ and $\mdust$ (the red solid line). A comparison of the solid red and blue dashed lines
indicates that incorporating the dust mass results in significantly more accurate SED predictions compared with using $\lir$ alone. The median
$\chi^2_r$ value is 0.56 (0.82) when both $\lir$ and $\mdust$ are (only $\lir$ is) used to predict C1. This result
indicates that the SEDs should be parameterized in terms of $\lir$ and $\mdust$, not just $\lir$; we shall discuss this in detail below.
The similarity of the green dashed and solid red lines indicates that using PC2 in addition to PC1 does not yield significantly better SED predictions
in terms of $\chi^2_r$; in this case, the median $\chi^2_r$ is 0.54, only 0.02 less than when PC1 alone is used (and both $\lir$ and $\mdust$ are
used to predict C1). This is consistent with our observation from Figure \ref{fig:reconstruct} that adding PC2 sometimes leads to a better
prediction at long wavelengths but sometimes causes the prediction to be worse.

\begin{figure}
\centering
\vskip -0.0cm
\resizebox{3.0in}{!}{\includegraphics[angle=0]{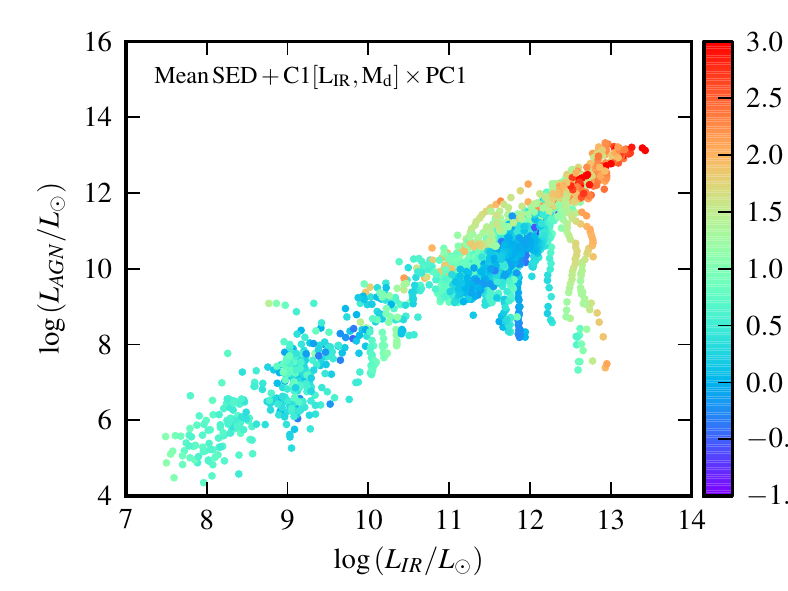}}
\resizebox{3.0in}{!}{\includegraphics[angle=0]{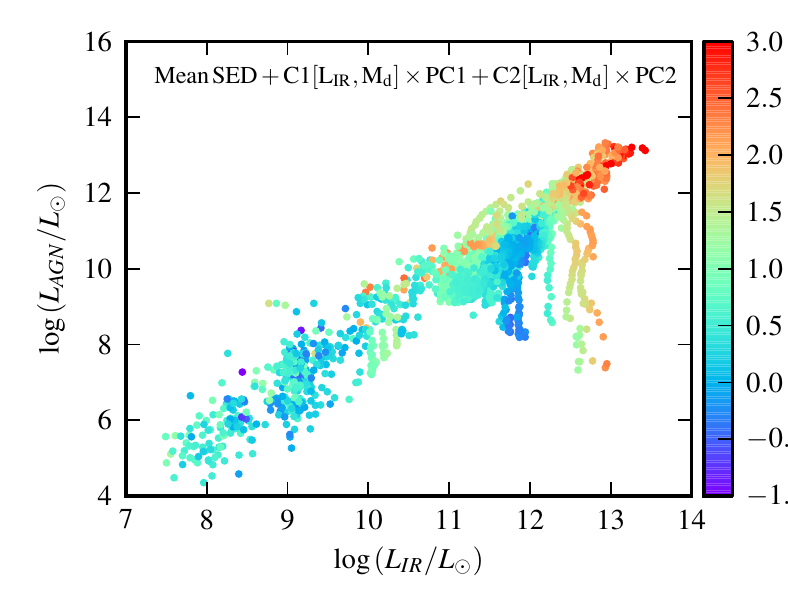}}
\caption{Shows the reduced $\chi^2$ values for the predicted SEDs. 
In the \emph{top} panel, the SEDs are predicted based on 
using only the first PC component, with its coefficient predicted using both IR luminosity and dust mass.
In the \emph{bottom} panel, the SEDs are predicted based on using both PC1 and PC2; for both PCs, the
coefficients are predicted using $\lir$ and dust mass, but the relations used differ (see Equations \ref{eq:c1}
and \ref{eq:c2}). The colors of the circles indicate the value of the logarithm of the reduced $\chi^2$.
The galaxies for which the SEDs are predicted least well
tend to be galaxies that have high AGN luminosities given their $\lir$ values
or/and have $\lir \ga 10^{12.5} \lsun$.}
\label{fig:chi2_plane}
\end{figure}

Figure~\ref{fig:compare_C1_C2} indicates that for a subset of galaxies, the SED prediction fails catastrophically ($\chi^2_r >> 10$).
By exploring the locations of these galaxies in different parameter spaces, we determined that the catastrophic failures tend
to have a high AGN contribution to the bolometric luminosity or/and $\lir \ga 10^{12.5} \lsun$, as indicated by Figure~\ref{fig:chi2_plane}.
This figure shows each galaxy in the plane of $\lir$ and $L_{\rm AGN}$.
The colors of the points indicate the $\chi^2$ value of the predicted SED. 
The top panel shows the results for the prediction based on only PC1, whereas the bottom panel shows the results for the 
prediction based on using both PC1 and PC2. Thus, the red circles represent galaxies for which the SED prediction fails
catastrophically. It is clear that in both cases, the galaxies for which the SEDs are predicted least well
tend to be galaxies that have high AGN luminosities given their $\lir$ values (i.e., high AGN luminosity fractions)
or/and have $\lir \ga 10^{12.5} \lsun$. We analyzed the
galaxies with high AGN fractions separately, but we were unable to determine global parameters that we could use to predict
their SEDs well. We speculate regarding the reasons for our inability to predict the FIR SEDs of such galaxies in Section
\ref{S:failures}. We retain the high-AGN galaxies in our subsequent analysis, but the results are not significantly changed if we exclude
them because only a small fraction of the simulated galaxies have $L_{\rm AGN} > 0.1$.

\section{Impact of dust mass on the SEDs} \label{sec:dust_mass}

\begin{figure}
\centering
\vskip -0.0cm
\resizebox{3.0in}{!}{\includegraphics[angle=0]{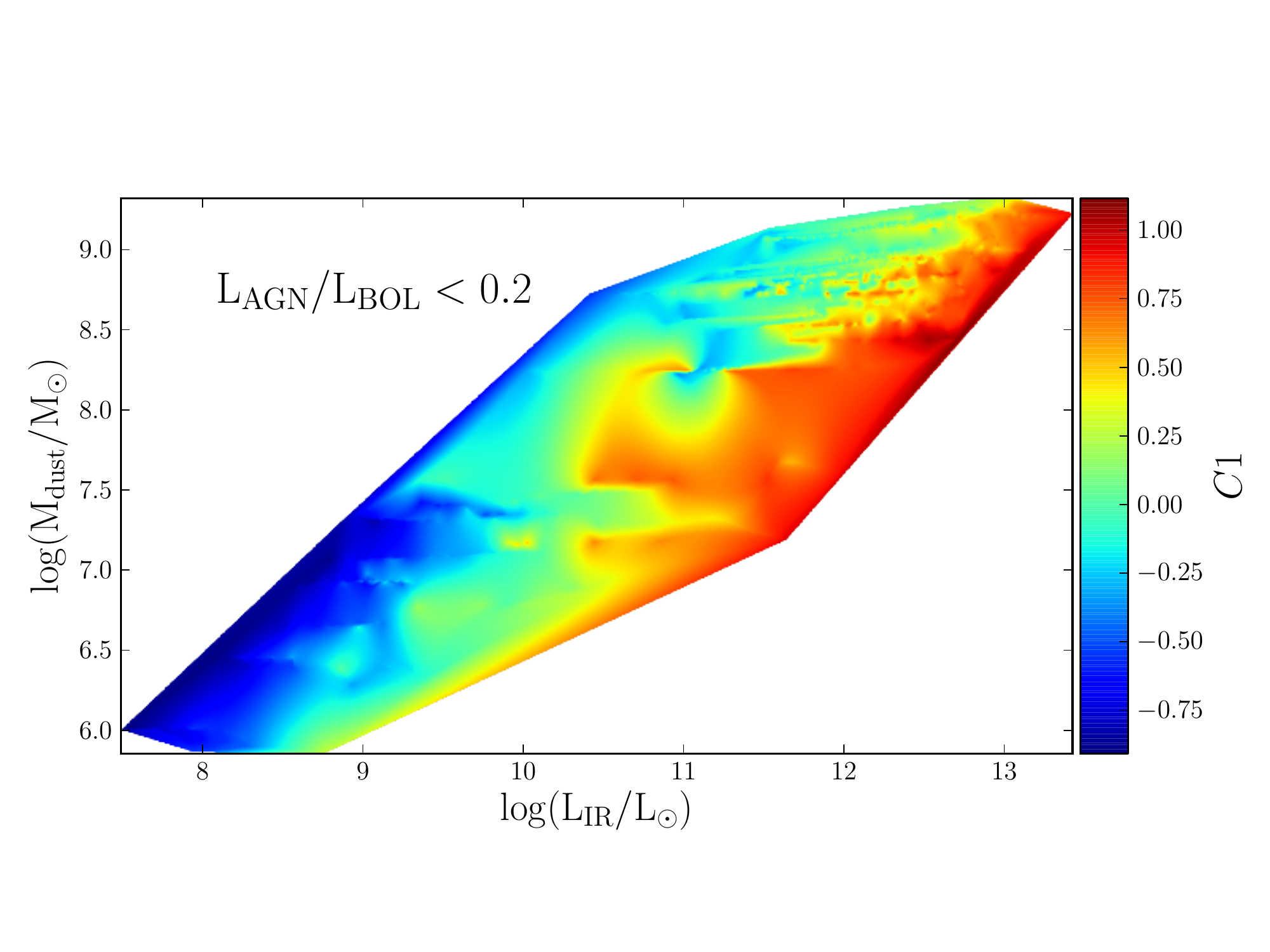}}
\caption{Shows the value of the C1 coefficient (see the colorbar) as a function of $\lir$(on the $y$-axis) and $\mdust$ (on the
$x$-axis). The interpolation between data points is performed using the nearest-neighbor method, 
which can result in some artifacts, but the trend that we see is robust. At fixed $\lir$, higher $\mdust$ results in
lower C1 values (i.e., lower effective dust temperature). At fixed $\mdust$, higher $\lir$ results in higher C1 values
(i.e., higher effective dust temperature).}
\label{fig:lir_mdust}
\end{figure}

FIR SEDs are often parameterized using templates that depend on the IR luminosity alone \citep[e.g.,][]{CE01,R09,2013Lee,2013Symeonidis}.
At fixed redshift, the effective dust temperature is observed to increase (i.e., the peak of the FIR SED shifts to shorter wavelengths) with increasing $\lir$
\citep[][and references therein]{casey14}.
As discussed above, Figure~\ref{fig:hist_compare} indicates that using both the IR luminosity and dust mass increases our ability to predict the C1 coefficients for 
galaxies compared with using the IR luminosity alone as a predictor. Because we know that C1 basically makes 
an SED cooler or warmer with respect to the mean SED (higher C1 values tend to make the SEDs hotter), it is instructive 
to examine how C1 behaves on the $\lir$ and $\mdust$ plane. This is shown in Figure~\ref{fig:lir_mdust}. 
One can see that at fixed $\lir$, higher values of $\mdust$ correspond to
lower values of C1, i.e., cooler SEDs. At fixed $\mdust$, higher values of $\lir$ lead to higher values of C1, i.e., hotter SEDs.

\begin{figure}
\centering
\vskip -0.0cm
\resizebox{3.0in}{!}{\includegraphics[angle=0]{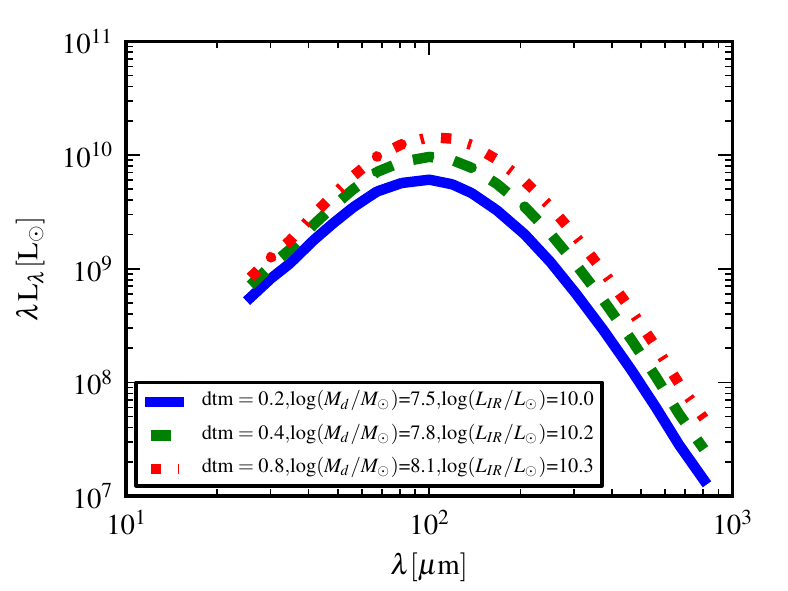}}
\resizebox{3.0in}{!}{\includegraphics[angle=0]{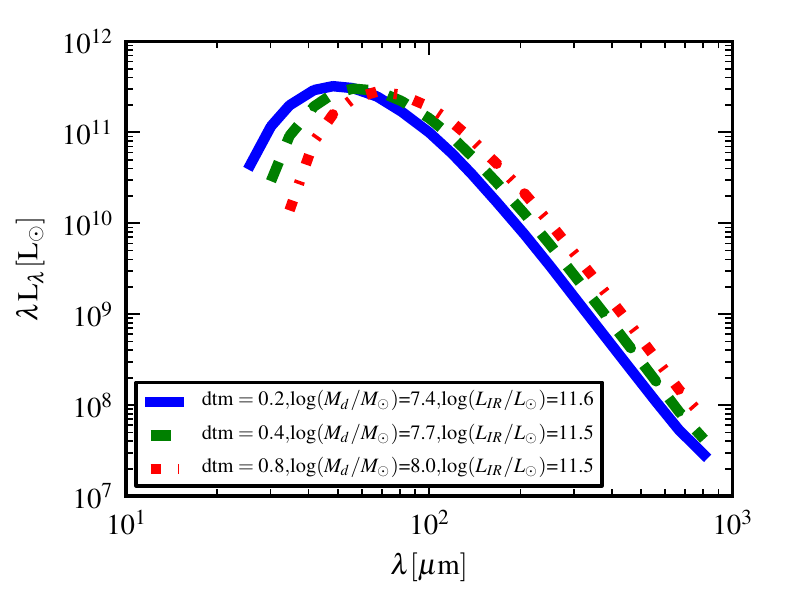}}
\caption{Shows the effect of changing the dust-to-metal ratio, and thus dust mass, on the SED. Each panel shows the SEDs of a simulated
galaxy at a single time and viewed from a fixed viewing angle for three different dust-to-metal ratios, as specified
in the legend; all other properties of the galaxies are unchanged.
The $\lir$ and $\mdust$ values for each SED are also shown in the legend. In the \emph{top} panel, the galaxy is not fully opaque when
the dust-to-metal ratio is 0.2. Consequently, as the dust-to-metal ratio is increased, $\lir$ increases by a factor of $\sim 2$,
and the SED shifts only slightly to the right. In the \emph{bottom} panel, $\lir$ decreases by 0.1 dex as the dust-to-metal ratio is increased;
this is a consequence of the viewing-angle dependence of $\lir$ (see the text for details). The SED systematically shifts
to longer wavelengths as the dust-to-metal ratio is increased, which is consistent with our expectation based on the relationship shown in
Figure~\ref{fig:lir_mdust} and considerations of thermal equilibrium.}
\label{fig:dtm}
\end{figure}

To more explicitly demonstrate the effect of the dust mass on the SEDs, we have
re-run a subset of the dust radiative transfer calculations with artificially lower 
and higher dust masses by changing the default dust-to-metal density ratio of 0.4
to 0.2 and 0.8, respectively. Consequently, the relative distribution of the dust
remains the same, but the overall normalization and
thus total dust masses are half or twice those of the standard run.
Representative results for two snapshots are presented in Figure~\ref{fig:dtm}; each
panel shows how the FIR SED for a single time and viewing angle varies as the
dust-to-metal density ratio is varied. The legends specify the assumed dust-to-metal ratio
and values of $\lir$ and $\mdust$ for each of the SEDs.

The results qualitatively agree with our expectations based on the above analysis.
The top panel corresponds to the pre-coalescence phase of the most-massive
major merger simulation from the $z = 0$ dataset. As the dust-to-metal ratio is
increased, $\lir$ increases, and the peak of the SED shifts slightly
to longer wavelengths (i.e., the SED becomes colder). In this case, a
substantial fraction of the luminosity is not absorbed when the
dust-to-metal ratio is 0.2. Consequently, increasing the dust-to-metal
ratio leads to higher optical depths and thus a larger fraction of
light being absorbed and re-emitted in the IR.

The SEDs shown in the bottom panel correspond to the M3M3e merger
simulation near the peak of the starburst induced at final
coalescence. Because the luminosity is dominated by a central
dust-enshrouded starburst, the simulated galaxy is already opaque to
effectively all of the radiation when the dust-to-metal ratio is
0.2. Consequently, $\lir$ does not increase as the dust-to-metal ratio
(and thus dust mass) is increased. Instead, it actually decreases by
0.1 dex; this occurs because we are considering the $\lir$ associated
with a single viewing angle. When there is a non-negligible optical
depth in the FIR, as can be the case for (U)LIRGs (see the discussion
in \citealt{H12}), $\lir$ can depend on the viewing angle. The
increased viewing-angle dependence as the dust-to-metal ratio is
increased explains the aforementioned decrease because as the
dust-to-metal ratio is increased, more of the short-wavelength IR
emission is removed from this particular line of sight. The absorbed
luminosity, which is independent of viewing angle, is effectively
identical in this case for all dust-to-metal ratios.

Because the $\lir$ values are almost identical for the three SEDs shown in the bottom panel, they provide
a clean test of how the SED varies with dust mass for fixed IR luminosity. We see that as
expected based on the trend shown in Figure~\ref{fig:lir_mdust}, the FIR SED systematically
shifts to longer wavelengths as the dust-to-metal ratio, and thus total dust mass, is increased;
all other properties of the galaxy are kept fixed.

\begin{figure}[!h]
\centering
\vskip -0.0cm
\resizebox{3.5in}{!}{\includegraphics[angle=0]{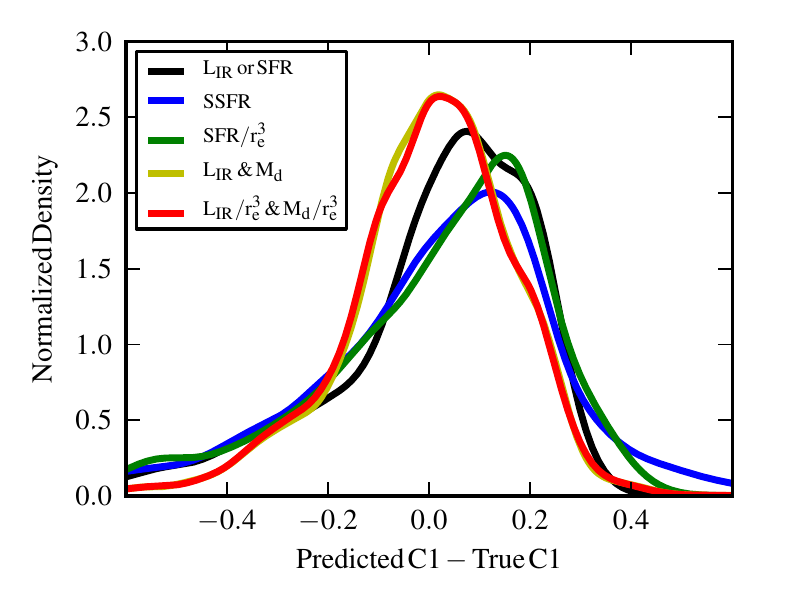}}
\resizebox{3.5in}{!}{\includegraphics[angle=0]{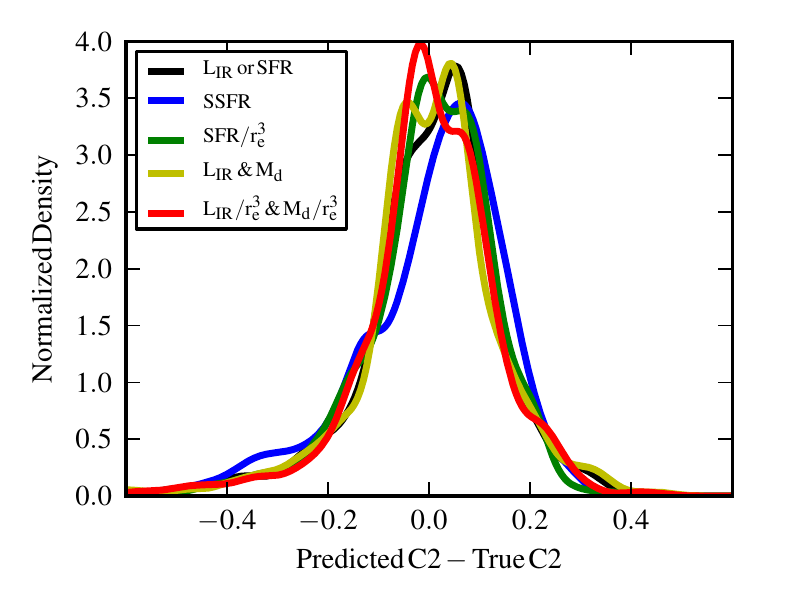}}
\caption{The difference between the predicted and true values of the coefficients for various estimators
(similar to Figure \ref{fig:compare_C1_C2}). In this case, some of the estimators used
include the galaxy sizes. The estimator used in each case is labeled in the legend, and in all cases, we use the logarithm
of the parameter. For both C1 (\emph{top}) and C2 (\emph{bottom}), incorporating the size does not significantly
increase the predictive power.}
\label{fig:C1_C2_SD_estimate}
\end{figure}

The reason why at fixed dust mass, increasing the IR luminosity increases the temperature
of the SED is that there are more photons available to heat the same amount of dust; consequently,
thermal equilibrium dictates that the dust temperature must increase.
The reason why at fixed IR luminosity, increasing the dust mass tends to make 
the SED cooler is that the luminosity is distributed over a greater mass of dust (because of dust self-absorption).
Thus, thermal equilibrium implies that the dust temperature will decrease. 
For an isothermal modified blackbody, the temperature scales as $T \propto (\lir/\mdust)^{1/(4+\beta)}$, where
$\beta$ is the power-law index of the dust opacity curve in the FIR \citep[e.g.][]{H11,Lanz14}.
Although the simulated SEDs are not quantitatively well-described by this simple model \citep{H11,H12,Lanz14},
it provides physical motivation for the claim that the effective dust temperature increases (decreases)
when $\lir$ ($\mdust$) is increased and $\mdust$ ($\lir$) is kept fixed.\footnote{For observational
values of $\lir$ and $\mdust$ inferred from fitting galaxy SEDs with an isothermal modified blackbody, this relation
is obeyed by construction. However, this is not the case for the simulations, in which the SED shape, and thus
effective dust temperature, can in principle depend not only on the total luminosity heating the dust and the
dust mass but also other factors, including the spatial distribution of the dust and sources and the dust composition.}
Put another way, the mean intensity `seen' by a dust grain is proportional to $\lir$/$\mdust$ (see \citealt{draine_li_07}
and \citealt{Draine2007} for detailed discussions). Thus, as $\lir$ is increased or $\mdust$ is decreased, the
mean intensity of light absorbed by the dust and thus typical grain temperature increases, and vice versa.

\section{Impact of galaxy sizes on the SED shape} \label{S:size_results}

In addition to the total absorbed luminosity and dust mass, both of which must affect the SED shape
because of thermal equilibrium, the geometry of radiation sources and dust can influence the SED.
The surface densities of various components (e.g., all stars, young stars, gas, and dust) are global
(observable) parameters that crudely characterize the global geometry of a galaxy and are often
used in observational studies to interpret the evolution in the FIR SED shapes of galaxies
\citep[e.g.,][]{Elbaz2011,2013Rujopakarn}.
Thus, it is worthwhile to investigate whether the PCA coefficients can be predicted better by
incorporating information regarding the sizes.

We calculated the 3D baryonic half-mass radii ($r_e$) and used these to approximate various
average volume densities by dividing integrated quantities -- such as the SFR, $\lir$, and stellar, gas, and dust
masses -- by the half-mass radii cubed.\footnote{We opted to use volume densities rather than surface
densities because the former are more relevant for the radiation transfer (albeit more difficult to infer
from observations).}
We then investigated whether various combinations of volume densities could be used to
predict the values of C1 and C2 better than using our standard parameterizations, C1($\lir$,$\mdust$)
and C2($\lir$,$\mdust$).

Figure~\ref{fig:C1_C2_SD_estimate} shows the difference between
the true values of the PCA coefficients and those predicted using various estimators, as indicated in 
the legends (for clarity, we show only a subset of the combinations that we tried; the others fared
comparably or worse). The top panel shows the results for C1, and the bottom shows those for C2.
To relate the coefficients' values to a single galaxy parameter, we used models of
the form $C1= A \log X + B$, where X is the galaxy parameter indicated in the figure legend and $A$
and $B$ are fitting coefficients. Similarly, when two parameters were used, we employed models of the form
$C1= A \log X + B \log Y + C$.

Examination of Figure~\ref{fig:C1_C2_SD_estimate} reveals that incorporating the size of the system
when predicting the coefficients yields yields at best a very marginal improvement in the
predictive power compared with using the total IR luminosity and dust mass values alone. Thus, we have
not used the sizes to predict the SEDs. This lack of improvement from incorporating the galaxy sizes
suggests that, at least for the simulated galaxies, the overall spatial extent does not play a significant
role in determining the shape of the FIR SED. We discuss this perhaps surprising result in detail in Section
\ref{S:size}.

\begin{figure*}
\centering
\vskip -0.0cm
\resizebox{7in}{!}{\includegraphics[angle=0]{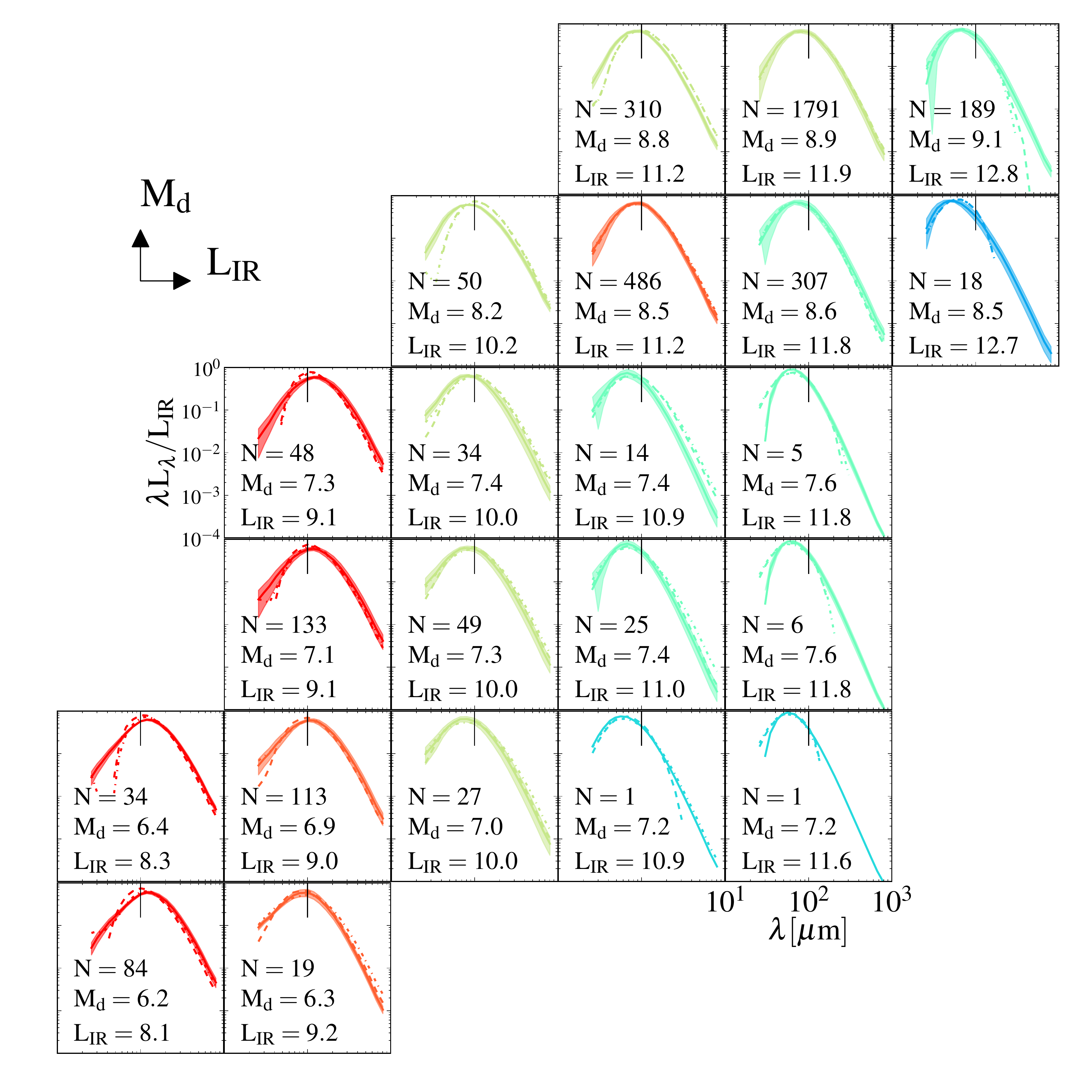}}
\caption{SEDs of our simulated galaxies binned according to IR luminosity and dust mass. In each panel, the median
FIR SED in that bin is indicated by the solid line, and the shaded region represents the 16 -- 84th percentile range of the FIR SEDs in that bin.
The dashed (dot-dashed) line shows the SED predicted using PC1 (PC1 and PC2), with the coefficient(s) predicted using the median $\lir$
and $\mdust$ values for that bin. The SEDs have been normalized by dividing by $\lir$.
The color coding is based on where the peak of the FIR SED is located, with redder colors corresponding to longer
peak wavelengths (i.e., colder effective dust temperatures).
The logarithms of the median dust mass and IR luminosity (in solar units) are indicated in each panel, and the number of SEDs
in each bin (N) is also shown.
The dust mass increases in the upward direction, and $\lir$ increases to the right. In most columns, the SEDs become redder
from bottom to top; this visually illustrates the trend for increasing $\mdust$ to result in cooler SEDs when $\lir$ is fixed. In a given row,
the SEDs become bluer from left to right because at fixed $\mdust$, the SEDs become hotter as $\lir$ is increased.
The trend with $\lir$ is more apparent that the with $\mdust$ because except for the bottom row, the $\lir$ values in a given row span 2-3 orders of
magnitude, whereas the $\mdust$ values in a given column span only 1-2 orders of magnitude.
Generally, the SEDs are predicted very well near their peaks. In some cases, the SEDs are underpredicted
significantly at long wavelengths, although use of PC2 reduces the amount by which the SED is underpredicted compared with using only PC1.}
\label{fig:2d_template_fir}
\end{figure*}

\begin{figure*}
\centering
\vskip -0.0cm
\resizebox{7in}{!}{\includegraphics[angle=0]{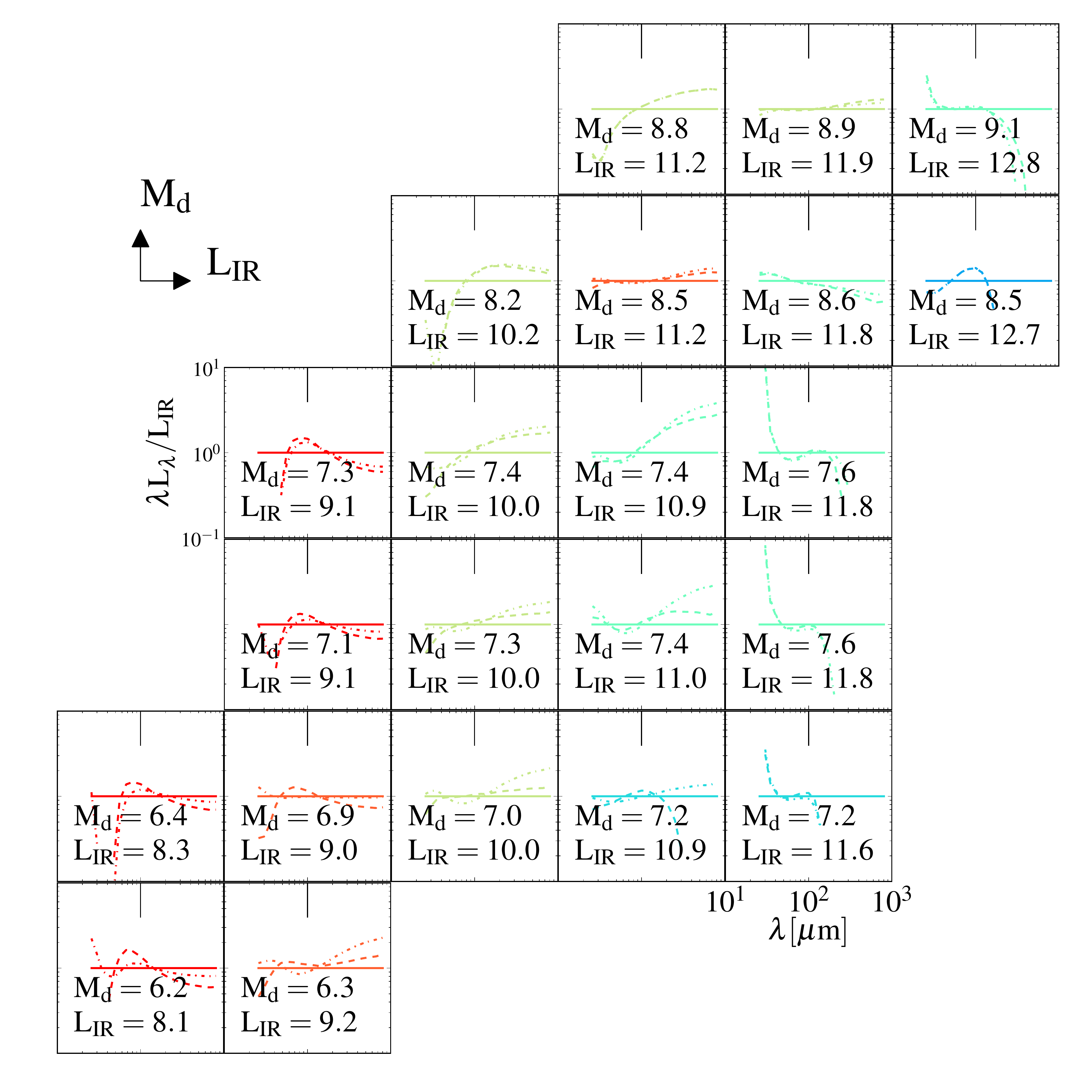}}
\caption{The predicted FIR SED divided by the true FIR SED when PC1 (PC1 and PC2) is (are) used to predict the median SED in each bin is shown
by the dashed (dot-dashed) lines. The solid lines correspond to a ratio of unity.
The $(\lir, \mdust)$ bins are the same as in Figure \ref{fig:2d_template_fir}.
This figure clearly shows that near their peaks, the median SEDs can be predicted well based on their $\lir$ and $\mdust$
values alone. At wavelengths significantly shorter or longer than the peak wavelength, the SEDs are predicted very well in some bins but are sometimes
incorrect by greater than an order of magnitude. Incorporating PC2 makes the predictions more accurate in some cases, but it can
also make them worse.}
\label{fig:2d_template_error}
\end{figure*}

\section{Two-parameter FIR SED templates} \label{S:templates}

Motivated by the results we have presented in the previous section, we now introduce a set of templates that is a function of both $\lir$ and $\mdust$.\footnote{The
templates are available at the following URL: \url{https://www.cfa.harvard.edu/~chayward/fir_sed_templates.html}.}
Motivated by the PCA results, we separated the simulations' FIR SEDs into $(\lir,\mdust)$ bins and calculated the median SED in each bin.
These are shown in Figure~\ref{fig:2d_template_fir}. Each panel shows the median FIR SED for each bin as a solid line, and the 16--84th percentile range
is indicated by the shaded area. The dashed lines represent the SEDs predicted by adding PC1 to the mean SED
with the coefficient value predicted by inputting the median $\lir$ and $\mdust$ values for each bin into Equation \ref{eq:c1}. The dot-dashed lines
indicate the SEDs predicted when both PC1 and PC2 are used. The SEDs have been
normalized by dividing by $\lir$ and are colored according to the wavelength at which the FIR peak is located, with redder colors indicating longer
wavelengths. The trends revealed in the above analysis are apparent from the templates: (1) at fixed $M_d$, the SEDs
become warmer as $\lir$ is increased (i.e., from left to right). (2) At fixed $\lir$, the SEDs become cooler when the $\mdust$ is increased (i.e., from bottom
to top).

Figure \ref{fig:2d_template_error} shows the ratios of the predicted SEDs to the
true SEDs, $L'_{\lambda}/L_{\lambda}$, both when only PC1 is used (solid lines) and when both PC1 and PC2 are used (dashed lines). These figures
demonstrate that the median SEDs in each bin are best predicted near their peaks. Shortward of the SED peaks, the median SEDs of most bins can be
predicted to within a few tens of percent. However, for some bins, the prediction is worse; in one (the bin with median values $\mdust = 10^{6.4} \msun$
and $\lir = 10^{8.3} \lsun$), the SED is underpredicted by more than an order of magnitude at $\lambda \sim 35$ \micron. At wavelengths $\ga 150$ \micron,
the SEDs are sometimes predicted to within a few tens of percent. However, in many bins, the SEDs are underpredicted considerably.
Using PC2 in addition to PC1 sometimes makes the prediction better (i.e., the dashed line is closer to 1 than is the solid line).

The goal of PCA is to explain the variance in a dataset. Thus, it is unsurprising that the peaks of the SEDs are predicted best, because this is the
region that dominates the variance in the dataset. Because the luminosity density at long wavelengths is $\sim 1-2$ orders of magnitude less
than that at the SED peak, the long-wavelength regions of the SEDs contribute relatively little to the variance. Consequently, it would be necessary
to use higher-order PCs to predict the long wavelength emission well.

\section{Discussion} \label{S:discussion}

\subsection{The simplicity of the FIR SEDs of galaxies} \label{S:simplicity}

We have found that the FIR SEDs of our simulated galaxies can be well predicted based on the galaxies' IR luminosities and dust masses
alone. Our results extend those of \citet{H11}, who demonstrated that the observed-frame submm flux densities of simulated $z \sim 2$
SMGs could be well predicted using these two parameters. Moreover, as noted in Section \ref{sec:PCA}, C1 effectively
depends on $\log \lir/\mdust$ because the best-fitting coefficients for ($\lir$ and $\mdust$) have almost the same magnitude
but opposite signs (0.52 and $-0.47$, respectively). Interestingly, in the simple isothermal modified blackbody model, the dust temperature
scales with $\lir/\mdust$.\footnote{However, it is important to note that how the effective dust temperatures of the simulated galaxies' SEDs depend
on $\lir/\mdust$ is not fully captured by this model \citep{Lanz14}.} In the more realistic case of a continuum of dust temperatures,
the mean intensity of the radiation absorbed by the dust is proportional to $\lir/\mdust$ \citep{draine_li_07}. Thus, both simple models and
our simulations suggest that the ratio $\lir/\mdust$ is a key determinant of the FIR SED of a galaxy.

These results indicate that the FIR SEDs of our simulated galaxies are rather simple. The skeptical reader may suggest that this simplicity
is a consequence of the simplicity of our simulations. However, even in the simulations, the FIR SEDs could in principle exhibit greater complexity
because the simulated galaxies contain dust with a continuum of temperatures (which are set by the 3D interstellar radiation field, 3D distribution
of dust, and grain properties). Thus, the fact that $\lir/\mdust$ encodes so much of the variance in the simulated SEDs is somewhat surprising.

Similar conclusions have been obtained based on observed FIR SEDs of galaxies. In particular, by fitting observed galaxy SEDs
using the model of \citet{draine_li_07}, \citet{Magdis2012} have argued
that the redshift evolution of `main sequence' galaxies' SEDs is driven by redshift evolution in the $\lir/\mdust$ ratio. For this reason, they
suggest using SED templates for `main sequence' galaxies that depend on $\lir/\mdust$.

\citet{Magnelli2014} studied how the effective temperature of galaxies depends on their position in the SFR--$M_{\star}$ plane and redshift.
They found that at all redshifts, the effective dust temperature smoothly increases with $\lir$, specific SFR, and the distance from the
`main sequence' (i.e. excess SSFR relative to what one would expect for a `main sequence' galaxy of the same mass), with the latter
two correlations being more significant than the first. They interpret the dependence on the distance from the `main sequence' in terms
of changes in the global star formation efficiency, SFR/$M_{\rm gas}$. However, we note that for an approximately constant dust ratio, this
quantity would serve as a proxy for $\lir/\mdust$. Thus, the results of \citet{Magnelli2014} are likely consistent with those of
\citet{Magdis2012} and this work.

\subsection{Observational support for the importance of dust mass}

Observations indicate that at fixed $\lir$, $z \sim 2-3$ galaxies have lower effective dust temperatures than do local galaxies
(\citealt{casey14} and references therein). Our results suggest that the observed trend could be a natural consequence of high-redshift
galaxies having more dust per unit IR luminosity. Thus, the evolution of the effective dust temperature--IR luminosity relation might be
a consequence of evolution in the global properties of the ISM of galaxies rather than changes in the small-scale geometry of star-forming regions.

The dust mass of a galaxy depends on the gas mass, gas-phase metallicity, and dust-to-metal ratio. As stars are formed, the ISM is enriched,
which can increase the dust mass. However, star formation simultaneously reduces the dust content of the ISM because the stars are formed
from dust-enriched gas. Simple models that encapsulate this competition between gas enrichment and consumption indicate that the maximum
dust mass of a galaxy depends weakly on the gas fraction: it is maximal when the gas fraction is 37 percent, but it varies by less than a factor
of 3 for gas fractions in the range of 4--86 percent \citep{Edmunds:1998}. This result is for a closed box, but the limit also holds if outflows
or unenriched inflows are allowed. Thus, although $z \sim 2-3$ ULIRGs are less metal-rich than local ULIRGs, they may still have higher dust
masses.

\begin{figure}
\centering
\vskip -0.0cm
\resizebox{3.5in}{!}{\includegraphics[angle=0]{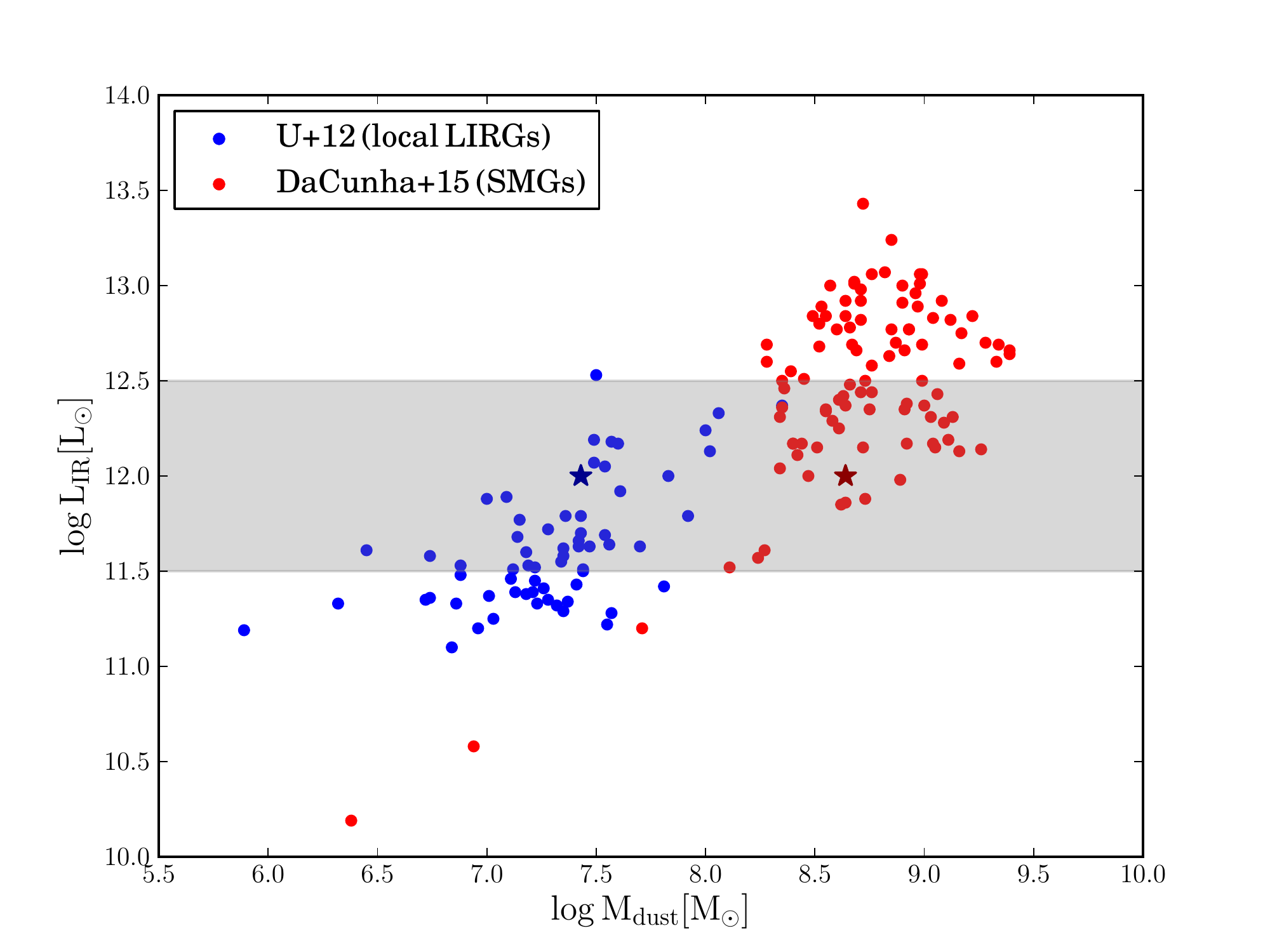}}
\caption{IR luminosity versus dust mass for two samples of IR-luminous galaxies, local LIRGs and ULIRGs from
\citet{2012U} and $z \sim 2-3$ SMGs from \citet{2015DaCunha}.
The stars indicate the median dust mass for the subset of each sample within the shaded region, which is defined
by $10^{11.5} < \lir/\lsun < 10^{12.5}$. The median dust mass of the SMGs is an order of magnitude greater
than that of the local (U)LIRGs.}
\label{fig:lirgs_vs_smgs}
\end{figure}

There is some observational evidence that supports our claim that the lower effective dust temperatures of $z \sim 2-3$ ULIRGs
are associated with higher dust masses relative to local ULIRGs.
Figure~\ref{fig:lirgs_vs_smgs} shows $\lir$ versus $\mdust$ for a sample of local LIRGs and ULIRGs \citep{2012U} and a sample of
$z \sim 2-3$ SMGs \citep{2015DaCunha}. For $\lir \sim 10^{11.5-12.5} \lsun$, the local ULIRGs have a median dust mass of
$10^{7.43} \msun$, whereas the $z \sim 2-3$ SMGs have a median $\mdust \approx 10^{8.64} \msun$. Thus, the SMGs have
dust masses that are more than an order of magnitude greater than those of the $z \sim 0$ ULIRGs. We argue that the increased
dust masses are the reason that the SMGs have cooler SEDs. This is consistent with the claim of \citet{Magdis2012}, who argued
that $z \sim 0.5-2$ `main sequence' ULIRGs have cooler SEDs than local ULIRGs because the former have lower $\lir/\mdust$ ratios.

We have used the $\lir$ and $\mdust$ data from \citet{2015DaCunha} because they were inferred from high-resolution ALMA data and \emph{Herschel}
data that were deblended based on the ALMA data. Thus, their photometry should be much less affected by blending than typical
datasets from \emph{Herschel} and other single-dish FIR/(sub)millimeter telescopes. This is highly desirable, because the fact
that blending becomes more severe at longer wavelengths could cause the SEDs extracted from blended data to be colder than the
true SEDs of individual sources. However, the significant caveat regarding our use of this dataset is that the SMG selection is biased
toward colder effective dust temperatures, and thus the $\lir/\mdust$ values of
$z \sim 2$ SMGs likely do not represent the full range of $\lir/\mdust$ values exhibited by $z \sim 2$ ULIRGs. To conclusively
determine how the $\lir/\mdust$ ratios of $z \sim 2$ ULIRGs compare with those of local ULIRGs, high-resolution FIR and (sub)millimeter
observations of a sufficiently large, unbiased (in terms of effective dust temperature) sample of $z \sim 2$ ULIRGs are
required.

\subsection{The unimportance of galaxy sizes in determining the SED shape} \label{S:size}

In Section \ref{S:size_results}, we demonstrated that incorporating information regarding the galaxy sizes did
not significantly improve our ability to predict the FIR SEDs. This result may be surprising to some readers,
given that it is often claimed that $z \sim 2$ galaxies have lower effective dust temperatures at fixed $\lir$
because they are more extended \citep[e.g.,][]{Elbaz2011,2013Rujopakarn}.
It is true that for a central source surrounded by dust (i.e., the
`shell' geometry of \citealt{2001Misselt}), increasing the spatial
extent of the absorbing material will result in colder dust because
the dust grains receive a more `diluted' radiation field. This
geometry may be a reasonable approximation (especially if the shell is
allowed to be clumpy) for individual HII regions and highly obscured
AGN. Indeed, more compact HII regions exhibit hotter dust
temperatures \citep[e.g.,][]{Groves:2008}. However, the overall
geometry of both real galaxies and our simulated galaxies is likely
more similar to the `dusty' geometry of \citet{2001Misselt}, in which
the stars and dust are mixed, because both the SFR density and dust
density are correlated with the gas density (see also
\citealt{Jonsson06b}). In such a geometry, the temperature of the
dust is insensitive to the size (assuming that the sizes of the
stellar and dust distributions are scaled in the same manner;
\citealt{2001Misselt}).

\subsection{The origin of catastrophic failures in the SED prediction} \label{S:failures}

Figure \ref{fig:chi2_plane} indicates that most galaxies for which the SED prediction is a catastrophic failure
(i.e., $\chi_r^2 >> 10$) have either high $L_{\rm AGN}/\lir$ or/and $\lir > 10^{12.5} \lsun$. There are a
few potential reasons that the SEDs of such sources would prove to be especially difficult to predict.
For the simulated galaxies in which the AGN contributes significantly to the bolometric luminosity, the AGN can
heat host-galaxy dust and cause FIR emission (this will be discussed in detail in Hayward et al., in prep.
and Roebuck et al., in prep.) Sources in which the AGN dominates the dust heating can differ from star
formation-dominated sources in terms of the SED of the radiation absorbed by the dust.
The geometry of such sources may also qualitatively differ: when the AGN dominates, the geometry is more
similar to the `shell' geometry of \citet{2001Misselt} than the mixed geometry, whereas the latter should
better describe sources in which star formation dominates the dust heating, as argued in the previous subsection.
Finally, in the AGN-dominated and most-IR luminous sources, dust self-absorption is likely more significant than in 
the other sources. Some or all of the above differences between star formation-powered and AGN-powered IR sources
may explain why the SEDs of many of the sources with high $L_{\rm AGN}/\lir$ or/and $\lir > 10^{12.5} \lsun$
cannot be predicted well.

\subsection{Implications for IR counts in hierarchical models}

Cosmological galaxy formation models have long struggled to correctly
reproduce the observed IR and sub-mm counts without introducing fairly
radical assumptions such as an extremely top-heavy stellar initial
mass function
(\citealt{Devriendt2000,Baugh2005,Lacey2010,Dave2010,2012Somerville,2012Niemi},
but cf. \citealt{H13}; see also the discussion in \citealt{casey14}). Due
to the infeasibility of carrying out full 3D radiative transfer
calculations on a cosmological hydrodynamic simulation (see
\ref{S:limitations}), to date most such calculations have relied on
semi-analytic models combined with a simplified approach to computing
the FIR SEDs. For example, the models of \citet{Devriendt2000} and
\citet{2012Somerville} used empirical libraries of dust emission
templates parameterized only by $\lir$. These libraries clearly cannot
capture the observed redshift evolution of the relationship between
effective dust temperature and IR luminosity. The SAMs presented by
\citet{2012Somerville} and further investigated by \citet{2012Niemi}
underpredicted IR number counts at wavelengths $\ga 100$ \micron, and
the discrepancy became worse with increasing wavelength. Our work here
suggests a straightforward and physically motivated way to improve the
modeling of dust emission in SAMs by using templates that depend on
both $\lir$ and $\mdust$. Our results are also encouraging for the use
of SAMs to model dust emission for large samples of galaxies, as they
suggest that the sizes and detailed geometries of galaxies (which are
properties that SAMs cannot model accurately) are sub-dominant
compared to the global parameters $\lir$ and $\mdust$.

If indeed the dust is colder (at fixed $\lir$) in high-redshift
galaxies, as indicated by observations, adopting two-parameter
templates like the ones we have presented here will clearly work in
the direction of alleviating the tension between the SAM predictions
and observations. Accounting for the effects of blending will further
reduce the discrepancy
\citep{HB13,H13,2015MNRAS.446.2291M,Cowley15}. However, whether
SAMs will predict a strong enough evolution of $\mdust$ with
$\lir$ and redshift to reproduce the observed IR-submm counts, once
observational effects such as blending have been taken into account,
remains to be seen. We plan to investigate this by incorporating
two-parameter dust emission templates in the \citet{2008Somerville}
SAM in a future work (Safarzadeh et al., in prep.).

\subsection{Limitations and future work} \label{S:limitations}

The detailed simulation methodology that we have employed in this work
has the advantages of being well studied,
and in previous works, the specific
simulations used in this work and similar simulations have been
demonstrated to reproduce the properties of a wide range of real
galaxies, as discussed in Section \ref{S:intro}. However, there are
naturally some limitations. First, because of the manner in which our
sample was constructed, the demographics of the population are by no
means representative of those of the real Universe. To achieve a
cosmologically representative population, it would be necessary to
perform 3D radiative transfer on galaxies selected from a large-volume
cosmological simulation. Unfortunately, such simulations typically
have spatial resolution $\ga 1$ kpc and thus do not resolve galaxy
disk scaleheights, much less the internal structure of the ISM.
Moreover, state-of-the-art cosmological simulations lack starbursts
(i.e., there are significantly fewer outliers above the `star formation
main sequence' than observed; \citealt{Sparre2015}), which are thought
to power local ULIRGs and a non-negligible fraction of $z \sim 2-3$
ULIRGs \citep{Hopkins:2010,HB13,H13,Cowley15}. Consequently, the
utility of such simulations for investigating the FIR SEDs of galaxies
remains limited. Cosmological `zoom-in' simulations can be used to
achieve orders-of-magnitude better spatial resolution; thus, radiative
transfer can be meaningfully applied to such simulations
\citep[e.g.,][]{Granato15}. However, the considerable computational
expense of such simulations strongly constrains the subset of the
parameter space that can be sampled.

Even for idealized, comparatively simple simulations such as ours, the
computational expense required to perform the radiative transfer is
significant. Consequently, the parameter space spanned by our
simulation suite is not exhaustive, and the sampling of the parameter
space is rather coarse. This limitation can be addressed in the future
through the use of a (considerably) larger simulation suite or through
performing radiative transfer on all resolved galaxies in a
large-volume cosmological simulation; however, we again stress that
for the latter, the limited spatial resolution will continue to be a
significant hurdle for the foreseeable future.
We have no reason to expect that our qualitative conclusions would
differ if we were to use a larger or/and cosmologically representative
simulation suite, especially given the demonstrated agreement between
our simulated galaxies' and real galaxies' SEDs. However, it is
possible that the details, such as the variation in the SED templates,
are sensitive to the specific simulations used.

Another significant limitation is that the hydrodynamical simulations
do not resolve the detailed structure of the ISM, both because of the
spatial resolution and the ISM model employed. As discussed in Section
\ref{sec:simulations}, we assume that the dust is uniformly
distributed on sub-resolution scales. Using our current methods, it is
possibly to crudely characterize the uncertainty associated with the
sub-resolution ISM structure by comparing two extremes, the default
model and one in which the dust in the cold clouds implicit in the
\citet{Springel03} ISM is completely ignored (i.e., the clumps have a
volume filling factor of zero). We have performed such a comparison
and found that the results do not qualitatively differ. In fact, it is
actually easier to predict the SEDs when the latter treatment is
used. We speculate that the reason for this result is that all optical
depths are smaller (by construction); consequently, dust self-absorption is less
significant.

Moreover, we by no means claim that our treatments of stellar and AGN feedback are state-of-the-art, and various groups are
now utilizing more sophisticated and likely more realistic feedback models \citep[e.g.,][]{Agertz2013,Agertz2015,Hopkins2014}.
However, to our knowledge, no UV--mm SEDs for such simulations have been presented in the literature; thus, the SEDs
used in this work still represent the state-of-the-art for UV--mm SEDs computed by performing dust radiative transfer on hydrodynamical
simulations. Computations of UV--mm SEDs of galaxies from the Feedback in Realistic Environments (FIRE) cosmological zoom-in
simulations \citep{Hopkins2014} using \emph{Sunrise} are underway, but this is a significant undertaking in and of itself. Our method
could be applied to these and other SED datasets in the future.

Finally, we have demonstrated that PCA is a useful tool for identifying which parameters drive the variation in galaxy SEDs.
However, the PCA results cannot be used to predict SEDs of galaxies outside of our parameter space because the mean SED
and PCs depend on the dataset on which the PCA is performed. Moreover, because of the dependence on the mean SED and
the fact that our simulation suite is not cosmologically representative in terms of the galaxy demographics, the PCA results
cannot even be used to predict the SEDs for samples in which the parameter space is a subset of that spanned by our simulations
but the distribution within the parameter space differs. Thus, the templates that we provide are a better tool for predicting FIR SEDs
than are the PCA results. Moreover, other statistical methods, such as neural networks, may prove to be more useful than PCA for this
purpose \citep[e.g.,][]{Silva2012}.

\section{Conclusions} \label{S:conclusions}

We performed PCA on a sample of FIR SEDs of simulated galaxies that we generated by performing dust radiative transfer
on hydrodynamical simulations in post-processing. Our goal was to determine what drives the variation in galaxies' FIR SEDs.
Our main conclusions are the following:

\begin{itemize}

\item The PCA indicated that only two PCs are sufficient to explain 97 percent of the variance in our SED sample. The first
component characterizes the peak of the SED, whereas the second characterizes the breadth of the peak.

\item The coefficient of the first PC, C1, is correlated with the IR luminosity, SFR, and AGN luminosity. This result indicates
that the SEDs are hotter when the IR luminosity, SFR, or AGN luminosity are greater.

\item Incorporating dust mass increases our ability to predict the value of C1 and thus the FIR SEDs. At fixed
IR luminosity, increased dust mass leads to lower C1 values and thus cooler SEDs.

\item The coefficient of the second PC, C2, is weakly anti-correlated with IR luminosity, SFR, AGN luminosity, and dust mass.
It can also be predicted using $\lir$ and $\mdust$, but the dependences on both quantities are weak.
Using the second PC improves how well the SEDs can be predicted in some cases but makes the predictions worse in others.

\item Examination of the catastrophic failures to reconstruct SEDs revealed that the bulk of such SEDs correspond to simulated
galaxies with high AGN fractions or/and $\lir > 10^{12.5} \lsun$.
For this sample, we were unable to predict the PC coefficients and thus SEDs well.

\item Incorporating galaxy sizes does not improve our ability to predict the SEDs.
 
\item The above conclusions suggest that the redshift evolution in effective dust temperature (i.e., at fixed $\lir$, $z \sim 2-3$ galaxies
exhibit lower effective dust temperatures compared with $z \sim 0$ galaxies) is not a consequence of higher-redshift ULIRGs being
more extended, as is often claimed. Instead, our work suggests that this difference is driven by $z \sim 2-3$ ULIRGs having higher
dust masses at fixed $\lir$ (because of their higher gas fractions than local galaxies), as suggested by some observations.

\item Because of the importance of dust mass in determining the FIR SED shape, a two parameter set of IR SED templates that depend
on both $\lir$ and $\mdust$ should be superior to those that depend on $\lir$ alone. We have generated such a set of templates
based on our simulated SEDs and made them publicly available. They should be useful for fitting observed galaxy SEDs and predicting
galaxy SEDs in unobserved wavelength regimes, and they can be used to predict IR SEDs of galaxies in cosmological simulations
and SAMs as long as the luminosity absorbed by dust and dust mass can be estimated.

\end{itemize}

\begin{acknowledgements}
We thank Maarten Baes for comments on the manuscript and Brice Menard
and Nick Scoville for useful discussions. CCH is grateful to the Gordon and Betty Moore
Foundation for financial support. CCH and RSS acknowledge the
hospitality of the Aspen Center for Physics, which is supported by the
National Science Foundation Grant No. PHY-1066293. RSS is grateful to
the Downsbrough family for their support and acknowledges support from
the Simons Foundation in the form of a Simons Investigator Award. This
work was partially supported by NASA's Astrophysics Data Analysis
Program, under grant NNX15AE54G.
\end{acknowledgements}

\bibliography{pca}

\begin{thebibliography}{}
\expandafter\ifx\csname natexlab\endcsname\relax\def\natexlab#1{#1}\fi

\bibitem[{{Agertz} \& {Kravtsov}(2015)}]{Agertz2015}
{Agertz}, O., \& {Kravtsov}, A.~V. 2015, \apj, 804, 18

\bibitem[{{Agertz} {et~al.}(2013){Agertz}, {Kravtsov}, {Leitner}, \&
  {Gnedin}}]{Agertz2013}
{Agertz}, O., {Kravtsov}, A.~V., {Leitner}, S.~N., \& {Gnedin}, N.~Y. 2013,
  \apj, 770, 25

\bibitem[{{Armus} {et~al.}(2009){Armus}, {Mazzarella}, {Evans}, {Surace},
  {Sanders}, {Iwasawa}, {Frayer}, {Howell}, {Chan}, {Petric}, {Vavilkin},
  {Kim}, {Haan}, {Inami}, {Murphy}, {Appleton}, {Barnes}, {Bothun}, {Bridge},
  {Charmandaris}, {Jensen}, {Kewley}, {Lord}, {Madore}, {Marshall},
  {Melbourne}, {Rich}, {Satyapal}, {Schulz}, {Spoon}, {Sturm}, {U}, {Veilleux},
  \& {Xu}}]{2009Armus}
{Armus}, L., {Mazzarella}, J.~M., {Evans}, A.~S., {et~al.} 2009, \pasp, 121,
  559

\bibitem[{{Baugh} {et~al.}(2005){Baugh}, {Lacey}, {Frenk}, {Granato}, {Silva},
  {Bressan}, {Benson}, \& {Cole}}]{Baugh2005}
{Baugh}, C.~M., {Lacey}, C.~G., {Frenk}, C.~S., {et~al.} 2005, \mnras, 356,
  1191

\bibitem[{{Bernhard} {et~al.}(2014){Bernhard}, {B{\'e}thermin}, {Sargent},
  {Buat}, {Mullaney}, {Pannella}, {Heinis}, \& {Daddi}}]{Bernhard2014}
{Bernhard}, E., {B{\'e}thermin}, M., {Sargent}, M., {et~al.} 2014, \mnras, 442,
  509

\bibitem[{{B{\'e}thermin} {et~al.}(2013){B{\'e}thermin}, {Wang}, {Dor{\'e}},
  {Lagache}, {Sargent}, {Daddi}, {Cousin}, \& {Aussel}}]{Bethermin2013}
{B{\'e}thermin}, M., {Wang}, L., {Dor{\'e}}, O., {et~al.} 2013, \aap, 557, A66

\bibitem[{{B{\'e}thermin} {et~al.}(2012){B{\'e}thermin}, {Daddi}, {Magdis},
  {Sargent}, {Hezaveh}, {Elbaz}, {Le Borgne}, {Mullaney}, {Pannella}, {Buat},
  {Charmandaris}, {Lagache}, \& {Scott}}]{Bethermin2012}
{B{\'e}thermin}, M., {Daddi}, E., {Magdis}, G., {et~al.} 2012, \apjl, 757, L23

\bibitem[{{Brassington} {et~al.}(2015){Brassington}, {Zezas}, {Ashby}, {Lanz},
  {Smith}, {Willner}, \& {Klein}}]{Brassington2015}
{Brassington}, N.~J., {Zezas}, A., {Ashby}, M.~L.~N., {et~al.} 2015, \apjs,
  218, 6

\bibitem[{{Casey} {et~al.}(2014){Casey}, {Narayanan}, \& {Cooray}}]{casey14}
{Casey}, C.~M., {Narayanan}, D., \& {Cooray}, A. 2014, \physrep, 541, 45

\bibitem[{Chakrabarti {et~al.}(2007)Chakrabarti, Cox, Hernquist, Hopkins,
  Robertson, \& {Di Matteo}}]{Chakrabarti:2007}
Chakrabarti, S., Cox, T.~J., Hernquist, L., {et~al.} 2007, \apj, 658, 840

\bibitem[{Chakrabarti {et~al.}(2008)Chakrabarti, Fenner, Cox, Hernquist, \&
  Whitney}]{Chakrabarti:2008}
Chakrabarti, S., Fenner, Y., Cox, T.~J., Hernquist, L., \& Whitney, B.~A. 2008,
  \apj, 688, 972

\bibitem[{Chakrabarti \& Whitney(2009)}]{Chakrabarti:2009}
Chakrabarti, S., \& Whitney, B.~A. 2009, \apj, 690, 1432

\bibitem[{{Chary} \& {Elbaz}(2001)}]{CE01}
{Chary}, R., \& {Elbaz}, D. 2001, \apj, 556, 562

\bibitem[{{Cowley} {et~al.}(2015){Cowley}, {Lacey}, {Baugh}, \&
  {Cole}}]{Cowley15}
{Cowley}, W.~I., {Lacey}, C.~G., {Baugh}, C.~M., \& {Cole}, S. 2015, \mnras,
  446, 1784

\bibitem[{{Cox} {et~al.}(2008){Cox}, {Jonsson}, {Somerville}, {Primack}, \&
  {Dekel}}]{Cox08}
{Cox}, T.~J., {Jonsson}, P., {Somerville}, R.~S., {Primack}, J.~R., \& {Dekel},
  A. 2008, \mnras, 384, 386

\bibitem[{{da Cunha} {et~al.}(2008){da Cunha}, {Charlot}, \& {Elbaz}}]{magphys}
{da Cunha}, E., {Charlot}, S., \& {Elbaz}, D. 2008, \mnras, 388, 1595

\bibitem[{{da Cunha} {et~al.}(2015){da Cunha}, {Walter}, {Smail}, {Swinbank},
  {Simpson}, {Decarli}, {Hodge}, {Weiss}, {van der Werf}, {Bertoldi},
  {Chapman}, {Cox}, {Danielson}, {Dannerbauer}, {Greve}, {Ivison}, {Karim}, \&
  {Thomson}}]{2015DaCunha}
{da Cunha}, E., {Walter}, F., {Smail}, I.~R., {et~al.} 2015, \apj, 806, 110

\bibitem[{{Dale} \& {Helou}(2002)}]{2002DH}
{Dale}, D.~A., \& {Helou}, G. 2002, \apj, 576, 159

\bibitem[{{Dale} {et~al.}(2001){Dale}, {Helou}, {Contursi}, {Silbermann}, \&
  {Kolhatkar}}]{2001Dale}
{Dale}, D.~A., {Helou}, G., {Contursi}, A., {Silbermann}, N.~A., \&
  {Kolhatkar}, S. 2001, \apj, 549, 215

\bibitem[{{Dav{\'e}} {et~al.}(2010){Dav{\'e}}, {Finlator}, {Oppenheimer},
  {Fardal}, {Katz}, {Kere{\v s}}, \& {Weinberg}}]{Dave2010}
{Dav{\'e}}, R., {Finlator}, K., {Oppenheimer}, B.~D., {et~al.} 2010, \mnras,
  404, 1355

\bibitem[{{De Geyter} {et~al.}(2014){De Geyter}, {Baes}, {Camps}, {Fritz}, {De
  Looze}, {Hughes}, {Viaene}, \& {Gentile}}]{DeGeyter2014}
{De Geyter}, G., {Baes}, M., {Camps}, P., {et~al.} 2014, \mnras, 441, 869

\bibitem[{{De Geyter} {et~al.}(2015){De Geyter}, {Baes}, {De Looze}, {Bendo},
  {Bourne}, {Camps}, {Cooray}, {De Zotti}, {Dunne}, {Dye}, {Eales}, {Fritz},
  {Furlanetto}, {Gentile}, {Hughes}, {Ivison}, {Maddox}, {Micha{\l}owski},
  {Smith}, {Valiante}, \& {Viaene}}]{DeGeyter2015}
{De Geyter}, G., {Baes}, M., {De Looze}, I., {et~al.} 2015, \mnras, 451, 1728

\bibitem[{{De Looze} {et~al.}(2012){De Looze}, {Baes}, {Fritz}, \&
  {Verstappen}}]{DeLooze2012}
{De Looze}, I., {Baes}, M., {Fritz}, J., \& {Verstappen}, J. 2012, \mnras, 419,
  895

\bibitem[{{De Looze} {et~al.}(2014){De Looze}, {Fritz}, {Baes}, {Bendo},
  {Cortese}, {Boquien}, {Boselli}, {Camps}, {Cooray}, {Cormier}, {Davies}, {De
  Geyter}, {Hughes}, {Jones}, {Karczewski}, {Lebouteiller}, {Lu}, {Madden},
  {R{\'e}my-Ruyer}, {Spinoglio}, {Smith}, {Viaene}, \& {Wilson}}]{DeLooze2014}
{De Looze}, I., {Fritz}, J., {Baes}, M., {et~al.} 2014, \aap, 571, A69

\bibitem[{{Desert} {et~al.}(1990){Desert}, {Boulanger}, \&
  {Puget}}]{Desert1990}
{Desert}, F.-X., {Boulanger}, F., \& {Puget}, J.~L. 1990, \aap, 237, 215

\bibitem[{{Devriendt} \& {Guiderdoni}(2000)}]{Devriendt2000}
{Devriendt}, J.~E.~G., \& {Guiderdoni}, B. 2000, \aap, 363, 851

\bibitem[{{Devriendt} {et~al.}(1999){Devriendt}, {Guiderdoni}, \&
  {Sadat}}]{Devriendt1999}
{Devriendt}, J.~E.~G., {Guiderdoni}, B., \& {Sadat}, R. 1999, \aap, 350, 381

\bibitem[{{Dom{\'{\i}}nguez-Tenreiro}
  {et~al.}(2014){Dom{\'{\i}}nguez-Tenreiro}, {Obreja}, {Granato}, {Schurer},
  {Alpresa}, {Silva}, {Brook}, \& {Serna}}]{GRASIL3D}
{Dom{\'{\i}}nguez-Tenreiro}, R., {Obreja}, A., {Granato}, G.~L., {et~al.} 2014,
  \mnras, 439, 3868

\bibitem[{{Dopita} {et~al.}(2005){Dopita}, {Groves}, {Fischera}, {Sutherland},
  {Tuffs}, {Popescu}, {Kewley}, {Reuland}, \& {Leitherer}}]{2005Dopita}
{Dopita}, M.~A., {Groves}, B.~A., {Fischera}, J., {et~al.} 2005, \apj, 619, 755

\bibitem[{{Draine} \& {Li}(2007)}]{draine_li_07}
{Draine}, B.~T., \& {Li}, A. 2007, \apj, 657, 810

\bibitem[{{Draine} {et~al.}(2007){Draine}, {Dale}, {Bendo}, {Gordon}, {Smith},
  {Armus}, {Engelbracht}, {Helou}, {Kennicutt}, {Li}, {Roussel}, {Walter},
  {Calzetti}, {Moustakas}, {Murphy}, {Rieke}, {Bot}, {Hollenbach}, {Sheth}, \&
  {Teplitz}}]{Draine2007}
{Draine}, B.~T., {Dale}, D.~A., {Bendo}, G., {et~al.} 2007, \apj, 663, 866

\bibitem[{Dwek(1998)}]{Dwek:1998}
Dwek, E. 1998, \apj, 501, 643

\bibitem[{Edmunds \& Eales(1998)}]{Edmunds:1998}
Edmunds, M.~G., \& Eales, S.~A. 1998, \mnras, 299, L29

\bibitem[{{Efstathiou} {et~al.}(2000){Efstathiou}, {Rowan-Robinson}, \&
  {Siebenmorgen}}]{2000Efstathiou}
{Efstathiou}, A., {Rowan-Robinson}, M., \& {Siebenmorgen}, R. 2000, \mnras,
  313, 734

\bibitem[{{Elbaz} {et~al.}(2011){Elbaz}, {Dickinson}, {Hwang},
  {D{\'{\i}}az-Santos}, {Magdis}, {Magnelli}, {Le Borgne}, {Galliano},
  {Pannella}, {Chanial}, {Armus}, {Charmandaris}, {Daddi}, {Aussel}, {Popesso},
  {Kartaltepe}, {Altieri}, {Valtchanov}, {Coia}, {Dannerbauer}, {Dasyra},
  {Leiton}, {Mazzarella}, {Alexander}, {Buat}, {Burgarella}, {Chary}, {Gilli},
  {Ivison}, {Juneau}, {Le Floc'h}, {Lutz}, {Morrison}, {Mullaney}, {Murphy},
  {Pope}, {Scott}, {Brodwin}, {Calzetti}, {Cesarsky}, {Charlot}, {Dole},
  {Eisenhardt}, {Ferguson}, {F{\"o}rster Schreiber}, {Frayer}, {Giavalisco},
  {Huynh}, {Koekemoer}, {Papovich}, {Reddy}, {Surace}, {Teplitz}, {Yun}, \&
  {Wilson}}]{Elbaz2011}
{Elbaz}, D., {Dickinson}, M., {Hwang}, H.~S., {et~al.} 2011, \aap, 533, A119

\bibitem[{{Fritz} {et~al.}(2006){Fritz}, {Franceschini}, \&
  {Hatziminaoglou}}]{Fritz2006}
{Fritz}, J., {Franceschini}, A., \& {Hatziminaoglou}, E. 2006, \mnras, 366, 767

\bibitem[{{Gonz{\'a}lez} {et~al.}(2011){Gonz{\'a}lez}, {Lacey}, {Baugh}, \&
  {Frenk}}]{Gonzalez:2011}
{Gonz{\'a}lez}, J.~E., {Lacey}, C.~G., {Baugh}, C.~M., \& {Frenk}, C.~S. 2011,
  \mnras, 413, 749

\bibitem[{{Gordon} {et~al.}(2001){Gordon}, {Misselt}, {Witt}, \&
  {Clayton}}]{2001Gordon}
{Gordon}, K.~D., {Misselt}, K.~A., {Witt}, A.~N., \& {Clayton}, G.~C. 2001,
  \apj, 551, 269

\bibitem[{{Granato} {et~al.}(2000){Granato}, {Lacey}, {Silva}, {Bressan},
  {Baugh}, {Cole}, \& {Frenk}}]{Granato:2000}
{Granato}, G.~L., {Lacey}, C.~G., {Silva}, L., {et~al.} 2000, \apj, 542, 710

\bibitem[{{Granato} {et~al.}(2015){Granato}, {Ragone-Figueroa},
  {Dom{\'{\i}}nguez-Tenreiro}, {Obreja}, {Borgani}, {De Lucia}, \&
  {Murante}}]{Granato15}
{Granato}, G.~L., {Ragone-Figueroa}, C., {Dom{\'{\i}}nguez-Tenreiro}, R.,
  {et~al.} 2015, \mnras, 450, 1320

\bibitem[{Groves {et~al.}(2008)Groves, Dopita, Sutherland, Kewley, Fischera,
  Leitherer, Brandl, \& van Breugel}]{Groves:2008}
Groves, B., Dopita, M.~A., Sutherland, R.~S., {et~al.} 2008, \apjs, 176, 438

\bibitem[{{Hayward} {et~al.}(2013{\natexlab{a}}){Hayward}, {Behroozi},
  {Somerville}, {Primack}, {Moreno}, \& {Wechsler}}]{HB13}
{Hayward}, C.~C., {Behroozi}, P.~S., {Somerville}, R.~S., {et~al.}
  2013{\natexlab{a}}, \mnras, 434, 2572

\bibitem[{{Hayward} {et~al.}(2012){Hayward}, {Jonsson}, {Kere{\v s}},
  {Magnelli}, {Hernquist}, \& {Cox}}]{H12}
{Hayward}, C.~C., {Jonsson}, P., {Kere{\v s}}, D., {et~al.} 2012, \mnras, 424,
  951

\bibitem[{{Hayward} {et~al.}(2011){Hayward}, {Kere{\v s}}, {Jonsson},
  {Narayanan}, {Cox}, \& {Hernquist}}]{H11}
{Hayward}, C.~C., {Kere{\v s}}, D., {Jonsson}, P., {et~al.} 2011, \apj, 743,
  159

\bibitem[{{Hayward} {et~al.}(2013{\natexlab{b}}){Hayward}, {Narayanan},
  {Kere{\v s}}, {Jonsson}, {Hopkins}, {Cox}, \& {Hernquist}}]{H13}
{Hayward}, C.~C., {Narayanan}, D., {Kere{\v s}}, D., {et~al.}
  2013{\natexlab{b}}, \mnras, 428, 2529

\bibitem[{{Hayward} \& {Smith}(2015)}]{HS15}
{Hayward}, C.~C., \& {Smith}, D.~J.~B. 2015, \mnras, 446, 1512

\bibitem[{{Hayward} {et~al.}(2014{\natexlab{a}}){Hayward}, {Torrey},
  {Springel}, {Hernquist}, \& {Vogelsberger}}]{H14arepo}
{Hayward}, C.~C., {Torrey}, P., {Springel}, V., {Hernquist}, L., \&
  {Vogelsberger}, M. 2014{\natexlab{a}}, \mnras, 442, 1992

\bibitem[{{Hayward} {et~al.}(2014{\natexlab{b}}){Hayward}, {Lanz}, {Ashby},
  {Fazio}, {Hernquist}, {Mart{\'{\i}}nez-Galarza}, {Noeske}, {Smith}, {Wuyts},
  \& {Zezas}}]{H14}
{Hayward}, C.~C., {Lanz}, L., {Ashby}, M.~L.~N., {et~al.} 2014{\natexlab{b}},
  \mnras, 445, 1598

\bibitem[{{Hopkins} {et~al.}(2014){Hopkins}, {Kere{\v s}}, {O{\~n}orbe},
  {Faucher-Gigu{\`e}re}, {Quataert}, {Murray}, \& {Bullock}}]{Hopkins2014}
{Hopkins}, P.~F., {Kere{\v s}}, D., {O{\~n}orbe}, J., {et~al.} 2014, \mnras,
  445, 581

\bibitem[{{Hopkins} {et~al.}(2010){Hopkins}, Younger, Hayward, Narayanan, \&
  Hernquist}]{Hopkins:2010}
{Hopkins}, P.~F., Younger, J.~D., Hayward, C.~C., Narayanan, D., \& Hernquist,
  L. 2010, \mnras, 402, 1693

\bibitem[{James {et~al.}(2002)James, Dunne, Eales, \& Edmunds}]{James:2002}
James, A., Dunne, L., Eales, S., \& Edmunds, M.~G. 2002, \mnras, 335, 753

\bibitem[{Jonsson(2006)}]{Jonsson06}
Jonsson, P. 2006, \mnras, 372, 2

\bibitem[{Jonsson {et~al.}(2006)Jonsson, Cox, Primack, \&
  Somerville}]{Jonsson06b}
Jonsson, P., Cox, T.~J., Primack, J.~R., \& Somerville, R.~S. 2006, \apj, 637,
  255

\bibitem[{Jonsson {et~al.}(2010)Jonsson, Groves, \& Cox}]{Jonsson10}
Jonsson, P., Groves, B.~A., \& Cox, T.~J. 2010, \mnras, 403, 17

\bibitem[{{Kelly} {et~al.}(2012){Kelly}, {Shetty}, {Stutz}, {Kauffmann},
  {Goodman}, \& {Launhardt}}]{Kelly:2012}
{Kelly}, B.~C., {Shetty}, R., {Stutz}, A.~M., {et~al.} 2012, \apj, 752, 55

\bibitem[{{Kennicutt} {et~al.}(2003){Kennicutt}, {Armus}, {Bendo}, {Calzetti},
  {Dale}, {Draine}, {Engelbracht}, {Gordon}, {Grauer}, {Helou}, {Hollenbach},
  {Jarrett}, {Kewley}, {Leitherer}, {Li}, {Malhotra}, {Regan}, {Rieke},
  {Rieke}, {Roussel}, {Smith}, {Thornley}, \& {Walter}}]{2003Kennicutt}
{Kennicutt}, Jr., R.~C., {Armus}, L., {Bendo}, G., {et~al.} 2003, \pasp, 115,
  928

\bibitem[{{Kov{\'a}cs} {et~al.}(2010){Kov{\'a}cs}, {Omont}, {Beelen},
  {Lonsdale}, {Polletta}, {Fiolet}, {Greve}, {Borys}, {Cox}, {De Breuck},
  {Dole}, {Dowell}, {Farrah}, {Lagache}, {Menten}, {Bell}, \&
  {Owen}}]{Kovacs:2010}
{Kov{\'a}cs}, A., {Omont}, A., {Beelen}, A., {et~al.} 2010, \apj, 717, 29

\bibitem[{{Lacey} {et~al.}(2010){Lacey}, {Baugh}, {Frenk}, {Benson}, {Orsi},
  {Silva}, {Granato}, \& {Bressan}}]{Lacey2010}
{Lacey}, C.~G., {Baugh}, C.~M., {Frenk}, C.~S., {et~al.} 2010, \mnras, 405, 2

\bibitem[{{Lanz} {et~al.}(2014){Lanz}, {Hayward}, {Zezas}, {Smith}, {Ashby},
  {Brassington}, {Fazio}, \& {Hernquist}}]{Lanz14}
{Lanz}, L., {Hayward}, C.~C., {Zezas}, A., {et~al.} 2014, \apj, 785, 39

\bibitem[{{Lanz} {et~al.}(2013){Lanz}, {Zezas}, {Brassington}, {Smith},
  {Ashby}, {da Cunha}, {Fazio}, {Hayward}, {Hernquist}, \&
  {Jonsson}}]{Lanz2013}
{Lanz}, L., {Zezas}, A., {Brassington}, N., {et~al.} 2013, \apj, 768, 90

\bibitem[{{Lee} {et~al.}(2013){Lee}, {Sanders}, {Casey}, {Scoville}, {Hung},
  {Le Floc'h}, {Ilbert}, {Aussel}, {Capak}, {Kartaltepe}, {Roseboom},
  {Salvato}, {Aravena}, {Berta}, {Bock}, {Oliver}, {Riguccini}, \&
  {Symeonidis}}]{2013Lee}
{Lee}, N., {Sanders}, D.~B., {Casey}, C.~M., {et~al.} 2013, \apj, 778, 131

\bibitem[{{Magdis} {et~al.}(2012){Magdis}, {Daddi}, {B{\'e}thermin}, {Sargent},
  {Elbaz}, {Pannella}, {Dickinson}, {Dannerbauer}, {da Cunha}, {Walter},
  {Rigopoulou}, {Charmandaris}, {Hwang}, \& {Kartaltepe}}]{Magdis2012}
{Magdis}, G.~E., {Daddi}, E., {B{\'e}thermin}, M., {et~al.} 2012, \apj, 760, 6

\bibitem[{{Magnelli} {et~al.}(2012){Magnelli}, {Lutz}, {Santini}, {Saintonge},
  {Berta}, {Albrecht}, {Altieri}, {Andreani}, {Aussel}, {Bertoldi},
  {B{\'e}thermin}, {Bongiovanni}, {Capak}, {Chapman}, {Cepa}, {Cimatti},
  {Cooray}, {Daddi}, {Danielson}, {Dannerbauer}, {Dunlop}, {Elbaz}, {Farrah},
  {F{\"o}rster Schreiber}, {Genzel}, {Hwang}, {Ibar}, {Ivison}, {Le Floc'h},
  {Magdis}, {Maiolino}, {Nordon}, {Oliver}, {P{\'e}rez Garc{\'{\i}}a},
  {Poglitsch}, {Popesso}, {Pozzi}, {Riguccini}, {Rodighiero}, {Rosario},
  {Roseboom}, {Salvato}, {Sanchez-Portal}, {Scott}, {Smail}, {Sturm},
  {Swinbank}, {Tacconi}, {Valtchanov}, {Wang}, \& {Wuyts}}]{Magnelli:2012}
{Magnelli}, B., {Lutz}, D., {Santini}, P., {et~al.} 2012, \aap, 539, A155

\bibitem[{{Magnelli} {et~al.}(2014){Magnelli}, {Lutz}, {Saintonge}, {Berta},
  {Santini}, {Symeonidis}, {Altieri}, {Andreani}, {Aussel}, {B{\'e}thermin},
  {Bock}, {Bongiovanni}, {Cepa}, {Cimatti}, {Conley}, {Daddi}, {Elbaz},
  {F{\"o}rster Schreiber}, {Genzel}, {Ivison}, {Le Floc'h}, {Magdis},
  {Maiolino}, {Nordon}, {Oliver}, {Page}, {P{\'e}rez Garc{\'{\i}}a},
  {Poglitsch}, {Popesso}, {Pozzi}, {Riguccini}, {Rodighiero}, {Rosario},
  {Roseboom}, {Sanchez-Portal}, {Scott}, {Sturm}, {Tacconi}, {Valtchanov},
  {Wang}, \& {Wuyts}}]{Magnelli2014}
{Magnelli}, B., {Lutz}, D., {Saintonge}, A., {et~al.} 2014, \aap, 561, A86

\bibitem[{{Mart{\'{\i}}nez-Galarza} {et~al.}(2014){Mart{\'{\i}}nez-Galarza},
  {Smith}, {Lanz}, {Hayward}, {Zezas}, {Rosenthal}, {Weiner}, {Hung}, {Ashby},
  \& {Groves}}]{Martinez14}
{Mart{\'{\i}}nez-Galarza}, J.~R., {Smith}, H.~A., {Lanz}, L., {et~al.} 2014,
  ArXiv e-prints, arXiv:1412.2760

\bibitem[{{Misselt} {et~al.}(2001){Misselt}, {Gordon}, {Clayton}, \&
  {Wolff}}]{2001Misselt}
{Misselt}, K.~A., {Gordon}, K.~D., {Clayton}, G.~C., \& {Wolff}, M.~J. 2001,
  \apj, 551, 277

\bibitem[{{Mu{\~n}oz Arancibia} {et~al.}(2015){Mu{\~n}oz Arancibia},
  {Navarrete}, {Padilla}, {Cora}, {Gawiser}, {Kurczynski}, \&
  {Ruiz}}]{2015MNRAS.446.2291M}
{Mu{\~n}oz Arancibia}, A.~M., {Navarrete}, F.~P., {Padilla}, N.~D., {et~al.}
  2015, \mnras, 446, 2291

\bibitem[{Narayanan {et~al.}(2010{\natexlab{a}})Narayanan, Hayward, Cox,
  Hernquist, Jonsson, Younger, \& Groves}]{N10}
Narayanan, D., Hayward, C.~C., Cox, T.~J., {et~al.} 2010{\natexlab{a}}, \mnras,
  401, 1613

\bibitem[{Narayanan {et~al.}(2010{\natexlab{b}})Narayanan, {Dey}, {Hayward},
  {Cox}, {Bussmann}, {Brodwin}, {Jonsson}, {Hopkins}, {Groves}, {Younger}, \&
  {Hernquist}}]{N10b}
Narayanan, D., {Dey}, A., {Hayward}, C.~C., {et~al.} 2010{\natexlab{b}},
  \mnras, 407, 1701

\bibitem[{{Niemi} {et~al.}(2012){Niemi}, {Somerville}, {Ferguson}, {Huang},
  {Lotz}, \& {Koekemoer}}]{2012Niemi}
{Niemi}, S.-M., {Somerville}, R.~S., {Ferguson}, H.~C., {et~al.} 2012, \mnras,
  421, 1539

\bibitem[{{Pilbratt} {et~al.}(2010){Pilbratt}, {Riedinger}, {Passvogel},
  {Crone}, {Doyle}, {Gageur}, {Heras}, {Jewell}, {Metcalfe}, {Ott}, \&
  {Schmidt}}]{Pilbratt2010}
{Pilbratt}, G.~L., {Riedinger}, J.~R., {Passvogel}, T., {et~al.} 2010, \aap,
  518, L1

\bibitem[{{Pope} {et~al.}(2008){Pope}, {Chary}, {Alexander}, {Armus},
  {Dickinson}, {Elbaz}, {Frayer}, {Scott}, \& {Teplitz}}]{Pope2008}
{Pope}, A., {Chary}, R.-R., {Alexander}, D.~M., {et~al.} 2008, \apj, 675, 1171

\bibitem[{{Popescu} {et~al.}(2011){Popescu}, {Tuffs}, {Dopita}, {Fischera},
  {Kylafis}, \& {Madore}}]{2011Popescu}
{Popescu}, C.~C., {Tuffs}, R.~J., {Dopita}, M.~A., {et~al.} 2011, \aap, 527,
  A109

\bibitem[{{Rieke} {et~al.}(2009){Rieke}, {Alonso-Herrero}, {Weiner},
  {P{\'e}rez-Gonz{\'a}lez}, {Blaylock}, {Donley}, \& {Marcillac}}]{R09}
{Rieke}, G.~H., {Alonso-Herrero}, A., {Weiner}, B.~J., {et~al.} 2009, \apj,
  692, 556

\bibitem[{{Robertson} {et~al.}(2006){Robertson}, Hernquist, Cox, {Di Matteo},
  Hopkins, Martini, \& Springel}]{Robertson06}
{Robertson}, B., Hernquist, L., Cox, T.~J., {et~al.} 2006, \apj, 641, 90

\bibitem[{{Rujopakarn} {et~al.}(2013){Rujopakarn}, {Rieke}, {Weiner},
  {P{\'e}rez-Gonz{\'a}lez}, {Rex}, {Walth}, \& {Kartaltepe}}]{2013Rujopakarn}
{Rujopakarn}, W., {Rieke}, G.~H., {Weiner}, B.~J., {et~al.} 2013, \apj, 767, 73

\bibitem[{{Siebenmorgen} \& {Kr{\"u}gel}(2007)}]{2007Siebenmorgen}
{Siebenmorgen}, R., \& {Kr{\"u}gel}, E. 2007, \aap, 461, 445

\bibitem[{{Silva} {et~al.}(2012){Silva}, {Fontanot}, \& {Granato}}]{Silva2012}
{Silva}, L., {Fontanot}, F., \& {Granato}, G.~L. 2012, \mnras, 423, 746

\bibitem[{{Silva} {et~al.}(1998){Silva}, {Granato}, {Bressan}, \&
  {Danese}}]{1998Silva}
{Silva}, L., {Granato}, G.~L., {Bressan}, A., \& {Danese}, L. 1998, \apj, 509,
  103

\bibitem[{{Smith} {et~al.}(2013){Smith}, {Hardcastle}, {Jarvis}, {Maddox},
  {Dunne}, {Bonfield}, {Eales}, {Serjeant}, {Thompson}, {Baes}, {Clements},
  {Cooray}, {De Zotti}, {Gonz{\`a}lez-Nuevo}, {Werf}, {Virdee}, {Bourne},
  {Dariush}, {Hopwood}, {Ibar}, \& {Valiante}}]{Smith:2013}
{Smith}, D.~J.~B., {Hardcastle}, M.~J., {Jarvis}, M.~J., {et~al.} 2013, \mnras,
  436, 2435

\bibitem[{{Snyder} {et~al.}(2011){Snyder}, {Cox}, {Hayward}, {Hernquist}, \&
  {Jonsson}}]{Snyder11}
{Snyder}, G.~F., {Cox}, T.~J., {Hayward}, C.~C., {Hernquist}, L., \& {Jonsson},
  P. 2011, \apj, 741, 77

\bibitem[{{Snyder} {et~al.}(2013){Snyder}, {Hayward}, {Sajina}, {Jonsson},
  {Cox}, {Hernquist}, {Hopkins}, \& {Yan}}]{Snyder:2013}
{Snyder}, G.~F., {Hayward}, C.~C., {Sajina}, A., {et~al.} 2013, \apj, 768, 168

\bibitem[{{Somerville} {et~al.}(2012){Somerville}, {Gilmore}, {Primack}, \&
  {Dom{\'{\i}}nguez}}]{2012Somerville}
{Somerville}, R.~S., {Gilmore}, R.~C., {Primack}, J.~R., \& {Dom{\'{\i}}nguez},
  A. 2012, \mnras, 423, 1992

\bibitem[{{Somerville} {et~al.}(2008){Somerville}, {Hopkins}, {Cox},
  {Robertson}, \& {Hernquist}}]{2008Somerville}
{Somerville}, R.~S., {Hopkins}, P.~F., {Cox}, T.~J., {Robertson}, B.~E., \&
  {Hernquist}, L. 2008, \mnras, 391, 481

\bibitem[{{Sparre} {et~al.}(2014){Sparre}, {Hartoog}, {Kr{\"u}hler}, {Fynbo},
  {Watson}, {Wiersema}, {D'Elia}, {Zafar}, {Afonso}, {Covino}, {de Ugarte
  Postigo}, {Flores}, {Goldoni}, {Greiner}, {Hjorth}, {Jakobsson}, {Kaper},
  {Klose}, {Levan}, {Malesani}, {Milvang-Jensen}, {Nardini}, {Piranomonte},
  {Sollerman}, {S{\'a}nchez-Ram{\'{\i}}rez}, {Schulze}, {Tanvir}, {Vergani}, \&
  {Wijers}}]{Sparre:2014}
{Sparre}, M., {Hartoog}, O.~E., {Kr{\"u}hler}, T., {et~al.} 2014, \apj, 785,
  150

\bibitem[{{Sparre} {et~al.}(2015){Sparre}, {Hayward}, {Springel},
  {Vogelsberger}, {Genel}, {Torrey}, {Nelson}, {Sijacki}, \&
  {Hernquist}}]{Sparre2015}
{Sparre}, M., {Hayward}, C.~C., {Springel}, V., {et~al.} 2015, \mnras, 447,
  3548

\bibitem[{Springel(2005)}]{Springel05gadget}
Springel, V. 2005, \mnras, 364, 1105

\bibitem[{Springel {et~al.}(2005)Springel, {Di Matteo}, \&
  Hernquist}]{Springel05feedback}
Springel, V., {Di Matteo}, T., \& Hernquist, L. 2005, \mnras, 361, 776

\bibitem[{Springel \& Hernquist(2003)}]{Springel03}
Springel, V., \& Hernquist, L. 2003, \mnras, 339, 289

\bibitem[{{Stalevski} {et~al.}(2012){Stalevski}, {Fritz}, {Baes}, {Nakos}, \&
  {Popovi{\'c}}}]{Stalevski2012}
{Stalevski}, M., {Fritz}, J., {Baes}, M., {Nakos}, T., \& {Popovi{\'c}}, L.~{\v
  C}. 2012, \mnras, 420, 2756

\bibitem[{{Symeonidis} {et~al.}(2013){Symeonidis}, {Vaccari}, {Berta}, {Page},
  {Lutz}, {Arumugam}, {Aussel}, {Bock}, {Boselli}, {Buat}, {Capak}, {Clements},
  {Conley}, {Conversi}, {Cooray}, {Dowell}, {Farrah}, {Franceschini},
  {Giovannoli}, {Glenn}, {Griffin}, {Hatziminaoglou}, {Hwang}, {Ibar},
  {Ilbert}, {Ivison}, {Floc'h}, {Lilly}, {Kartaltepe}, {Magnelli}, {Magdis},
  {Marchetti}, {Nguyen}, {Nordon}, {O'Halloran}, {Oliver}, {Omont},
  {Papageorgiou}, {Patel}, {Pearson}, {P{\'e}rez-Fournon}, {Pohlen}, {Popesso},
  {Pozzi}, {Rigopoulou}, {Riguccini}, {Rosario}, {Roseboom}, {Rowan-Robinson},
  {Salvato}, {Schulz}, {Scott}, {Seymour}, {Shupe}, {Smith}, {Valtchanov},
  {Wang}, {Xu}, {Zemcov}, \& {Wuyts}}]{2013Symeonidis}
{Symeonidis}, M., {Vaccari}, M., {Berta}, S., {et~al.} 2013, \mnras, 431, 2317

\bibitem[{{Takagi} {et~al.}(2003){Takagi}, {Arimoto}, \& {Hanami}}]{2003Takagi}
{Takagi}, T., {Arimoto}, N., \& {Hanami}, H. 2003, \mnras, 340, 813

\bibitem[{{U} {et~al.}(2012){U}, {Sanders}, {Mazzarella}, {Evans}, {Howell},
  {Surace}, {Armus}, {Iwasawa}, {Kim}, {Casey}, {Vavilkin}, {Dufault},
  {Larson}, {Barnes}, {Chan}, {Frayer}, {Haan}, {Inami}, {Ishida},
  {Kartaltepe}, {Melbourne}, \& {Petric}}]{2012U}
{U}, V., {Sanders}, D.~B., {Mazzarella}, J.~M., {et~al.} 2012, \apjs, 203, 9

\bibitem[{{Witt} \& {Gordon}(1996)}]{1996aWitt}
{Witt}, A.~N., \& {Gordon}, K.~D. 1996, \apj, 463, 681

\bibitem[{{Witt} \& {Gordon}(2000)}]{2000bWitt}
---. 2000, \apj, 528, 799

\bibitem[{Wuyts {et~al.}(2010)Wuyts, Cox, Hayward, Franx, Hernquist, Hopkins,
  Jonsson, \& van Dokkum}]{Wuyts10}
Wuyts, S., Cox, T.~J., Hayward, C.~C., {et~al.} 2010, \apj, 722, 1666

\bibitem[{{Younger} {et~al.}(2009){Younger}, Hayward, Narayanan, Cox,
  Hernquist, \& Jonsson}]{Younger09}
{Younger}, J.~D., Hayward, C.~C., Narayanan, D., {et~al.} 2009, \mnras, 396,
  L66

\end{thebibliography}

\end{document}